\documentclass[useAMS,usenatbib]{mn2e}
\usepackage{graphicx,natbib,times,amssymb}
\usepackage[T1]{fontenc} 
 \usepackage{aecompl} 
\title[Galaxy Zoo: The Green Valley is a Red Herring]{The Green Valley is a Red Herring: \\
Galaxy Zoo reveals two evolutionary pathways towards quenching of star formation in early- and late-type galaxies\thanks{This publication has been made possible by the participation of more than 250,000 volunteers in the Galaxy Zoo project. Their contributions are individually acknowledged at \texttt{http://www.galaxyzoo.org/Volunteers.aspx}.}}
\author[Kevin Schawinski et al.]{
 \parbox[t]{18cm}{
 Kevin Schawinski$^{1}$\thanks{E-mail: kevin.schawinski@phys.ethz.ch, Twitter: @kevinschawinski}, 
 C. Megan Urry$^{2,3,4}$, 
 Brooke D. Simmons$^{5}$, 
 Lucy Fortson$^{6}$,
 Sugata Kaviraj$^{7}$, 
 William C. Keel$^{8}$,
Chris J. Lintott$^{5,9}$, 
Karen L. Masters$^{10,11}$,
Robert C. Nichol$^{10,11}$, 
Marc Sarzi$^{7}$, 
Ramin, Skibba$^{12}$, 
Ezequiel Treister$^{13}$, 
Kyle W. Willett$^{6}$, 
O. Ivy Wong$^{14}$ and
Sukyoung K. Yi$^{15}$\\
 }\\
$^{1}$Institute for Astronomy, Department of Physics, ETH Zurich, Wolfgang-Pauli-Strasse 27, CH-8093 Zurich, Switzerland\\
$^{2}$ Department of Physics, Yale University, P.O. Box 208120, New Haven, CT 06520-8120, USA\\
$^{3}$ Yale Center for Astronomy and Astrophysics, Yale University, PO Box 208121, New Haven, CT 06520, USA\\
$^{4}$ Department of Astronomy, Yale University, P.O. Box 208101, New Haven, CT 06520-8101, USA\\
$^{5}$ Oxford Astrophysics, Denys Wilkinson Building, Keble Road, Oxford OX1 3RH, UK\\
$^{6}$ School of Physics and Astronomy, University of Minnesota, Minneapolis, MN 55455, USA\\
$^{7}$ Centre for Astrophysics Research, University of Hertfordshire, Hatfield, Herts AL1 9AB, UK\\
$^{8}$ Department of Physics and Astronomy, University of Alabama, Box 870324, Tuscaloosa, AL 35487, USA\\
$^{9}$ Adler Planetarium, 1300 S. Lakeshore Drive, Chicago, IL 60605\\
$^{10}$ Institute of Cosmology and Gravitation, University of Portsmouth, Dennis Sciama Building, Burnaby Road, Portsmouth, PO1 3FX, UK\\
$^{11}$ South East Physics Network; SEPNet; www.sepnet.ac.uk\\
$^{12}$ Center for Astrophysics and Space Sciences, Department of Physics, University of California, 9500 Gilman Drive, San Diego, CA 92093, USA\\
$^{13}$ Universidad de Concepci\'{o}n, Departamento de Astronom\'{\i}a, Casilla 160-C, Concepci\'{o}n, Chile\\
$^{14}$ CSIRO Astronomy \& Space Science, PO Box 76, Epping, NSW 1710, Australia\\
$^{15}$Department of Astronomy and Yonsei University Observatory, Yonsei University, Seoul 120-749, Republic of Korea
}

\begin{document}

\newcommand\aj{{AJ}}%
 % Astronomical Journal
\newcommand\actaa{{Acta Astron.}}%
 % Acta Astronomica
\newcommand\araa{{ARA\&A}}%
 % Annual Review of Astron and Astrophys
\newcommand\apj{{ApJ}}%
 % Astrophysical Journal
\newcommand\apjl{{ApJ}}%
 % Astrophysical Journal, Letters
\newcommand\apjs{{ApJS}}%
 % Astrophysical Journal, Supplement
\newcommand\ao{{Appl.~Opt.}}%
 % Applied Optics
\newcommand\apss{{Ap\&SS}}%
 % Astrophysics and Space Science
\newcommand\aap{{A\&A}}%
 % Astronomy and Astrophysics
\newcommand\aapr{{A\&A~Rev.}}%
 % Astronomy and Astrophysics Reviews
\newcommand\aaps{{A\&AS}}%
 % Astronomy and Astrophysics, Supplement
\newcommand\azh{{AZh}}%
 % Astronomicheskii Zhurnal
\newcommand\baas{{BAAS}}%
 % Bulletin of the AAS
\newcommand\caa{{Chinese Astron. Astrophys.}}%
 % Chinese Astronomy and Astrophysics
\newcommand\cjaa{{Chinese J. Astron. Astrophys.}}%
 % Chinese Journal of Astronomy and Astrophysics
\newcommand\icarus{{Icarus}}%
 % Icarus
\newcommand\jcap{{J. Cosmology Astropart. Phys.}}%
 % Journal of Cosmology and Astroparticle Physics
\newcommand\jrasc{{JRASC}}%
 % Journal of the RAS of Canada
\newcommand\memras{{MmRAS}}%
 % Memoirs of the RAS
\newcommand\mnras{{MNRAS}}%
 % Monthly Notices of the RAS
\newcommand\na{{New A}}%
 % New Astronomy
\newcommand\nar{{New A Rev.}}%
 % New Astronomy Review
\newcommand\pra{{Phys.~Rev.~A}}%
 % Physical Review A: General Physics
\newcommand\prb{{Phys.~Rev.~B}}%
 % Physical Review B: Solid State
\newcommand\prc{{Phys.~Rev.~C}}%
 % Physical Review C
\newcommand\prd{{Phys.~Rev.~D}}%
 % Physical Review D
\newcommand\pre{{Phys.~Rev.~E}}%
 % Physical Review E
\newcommand\prl{{Phys.~Rev.~Lett.}}%
 % Physical Review Letters
\newcommand\pasa{{PASA}}%
 % Publications of the Astron. Soc. of Australia
\newcommand\pasp{{PASP}}%
 % Publications of the ASP
\newcommand\pasj{{PASJ}}%
 % Publications of the ASJ
\newcommand\qjras{{QJRAS}}%
 % Quarterly Journal of the RAS
\newcommand\rmxaa{{Rev. Mexicana Astron. Astrofis.}}%
 % Revista Mexicana de Astronomia y Astrofisica
\newcommand\skytel{{S\&T}}%
 % Sky and Telescope
\newcommand\solphys{{Sol.~Phys.}}%
 % Solar Physics
\newcommand\sovast{{Soviet~Ast.}}%
 % Soviet Astronomy
\newcommand\ssr{{Space~Sci.~Rev.}}%
 % Space Science Reviews
\newcommand\zap{{ZAp}}%
 % Zeitschrift fuer Astrophysik
\newcommand\nat{{Nature}}%
 % Nature
\newcommand\iaucirc{{IAU~Circ.}}%
 % IAU Cirulars
\newcommand\aplett{{Astrophys.~Lett.}}%
 % Astrophysics Letters and Communications
\newcommand\apspr{{Astrophys.~Space~Phys.~Res.}}%
 % Astrophysics Space Physics Research
\newcommand\bain{{Bull.~Astron.~Inst.~Netherlands}}%
 % Bulletin Astronomical Institute of the Netherlands
\newcommand\fcp{{Fund.~Cosmic~Phys.}}%
 % Fundamental Cosmic Physics
\newcommand\gca{{Geochim.~Cosmochim.~Acta}}%
 % Geochimica Cosmochimica Acta
\newcommand\grl{{Geophys.~Res.~Lett.}}%
 % Geophysics Research Letters
\newcommand\jcp{{J.~Chem.~Phys.}}%
 % Journal of Chemical Physics
\newcommand\jgr{{J.~Geophys.~Res.}}%
 % Journal of Geophysical Research
\newcommand\jqsrt{{J.~Quant.~Spec.~Radiat.~Transf.}}%
 % Journal of Quantitiative Spectroscopy and Radiative Trasfer
\newcommand\memsai{{Mem.~Soc.~Astron.~Italiana}}%
 % Mem. Societa Astronomica Italiana
\newcommand\nphysa{{Nucl.~Phys.~A}}%
 % Nuclear Physics A
\newcommand\physrep{{Phys.~Rep.}}%
 % Physics Reports
\newcommand\physscr{{Phys.~Scr}}%
 % Physica Scripta
\newcommand\planss{{Planet.~Space~Sci.}}%
 % Planetary Space Science
\newcommand\procspie{{Proc.~SPIE}}%
 % Proceedings of the SPIE
\newcommand\helvet{{Helvetica~Phys.~Acta}}%
 % Helvetica Phys. Acta?

% Common satellites
\def\Chandra{\textit{Chandra}}
\def\XMM{\textit{XMM-Newton}}
\def\Swift{\textit{Swift}}

% Forbidden Lines
\def\OI{[\mbox{O\,{\sc i}}]~$\lambda 6300$}
\def\OIII{[\mbox{O\,{\sc iii}}]~$\lambda 5007$}
\def\SII{[\mbox{S\,{\sc ii}}]~$\lambda \lambda 6717,6731$}
\def\NII{[\mbox{N\,{\sc ii}}]~$\lambda 6584$}

% Balmer lines
\def\Ha{{H$\alpha$}}
\def\Hb{{H$\beta$}}

% Line ratios
\def\NIIHa{[\mbox{N\,{\sc ii}}]/H$\alpha$}
\def\SIIHa{[\mbox{S\,{\sc ii}}]/H$\alpha$}
\def\OIHa{[\mbox{O\,{\sc i}}]/H$\alpha$}
\def\OIIIHb{[\mbox{O\,{\sc iii}}]/H$\beta$}

% Common terms
\def\Ebmv{E($B-V$)}
\def\LOIII{$L[\mbox{O\,{\sc iii}}]$}
\def\Ledd{${L/L_{\rm Edd}}$}
\def\LOIIIs4{$L[\mbox{O\,{\sc iii}}]$/$\sigma^4$}
\def\LOIIIMbh{$L[\mbox{O\,{\sc iii}}]$/$M_{\rm BH}$}
\def\Mbh{$M_{\rm BH}$}
\def\Msigma{$M_{\rm BH} --- \sigma$}
\def\Ms{$M_{\rm *}$}
\def\Msun{$M_{\odot}$}
\def\Msunyr{$M_{\odot}yr^{-1}$}

% Units
\def\ergs{$~\rm erg~s^{-1}$}
\def\kms{$~\rm km~s^{-1}$}

% Software
\def\galfit{\texttt{GALFIT}}
\def\multidrizzle{\texttt{multidrizzle}}

% Other
\def\sersic{S\'{e}rsic}

\date{}

\pagerange{\pageref{firstpage}--\pageref{lastpage}} \pubyear{2013}

\maketitle

\label{firstpage}

\begin{abstract}
We use SDSS+\textit{GALEX}+Galaxy Zoo data to study the quenching of star formation in low-redshift galaxies. We show that the green valley between the blue cloud of star-forming galaxies and the red sequence of quiescent galaxies in the colour-mass diagram is not a single transitional state through which most blue galaxies evolve into red galaxies. Rather, an analysis that takes morphology into account makes clear that only a small population of blue early-type galaxies move rapidly across the green valley after the morphologies are transformed from disk to spheroid and star formation is quenched rapidly. In contrast, the majority of blue star-forming galaxies have significant disks, and they retain their late-type morphologies as their star formation rates decline very slowly. We summarize a range of observations that lead to these conclusions, including UV-optical colours and halo masses, which both show a striking dependence on morphological type. We interpret these results in terms of the evolution of cosmic gas supply and gas reservoirs. We conclude that late-type galaxies are consistent with a scenario where the cosmic supply of gas is shut off, perhaps at a critical halo mass, followed by a slow exhaustion of the remaining gas over several Gyr, driven by secular and/or environmental processes. In contrast, early-type galaxies require a scenario where the gas supply and gas reservoir are destroyed virtually instantaneously, with rapid quenching accompanied by a morphological transformation from disk to spheroid. This gas reservoir destruction could be the consequence of a major merger, which in most cases transforms galaxies from disk to elliptical morphology, and mergers could play a role in inducing black hole accretion and possibly AGN feedback.
\end{abstract}

%-----------------------------------------------------------------------------------------------------------------------------------
\section{Introduction}
\label{sec:intro}

\begin{keywords}
galaxies: evolution; galaxies: active; galaxies: spiral; galaxies: elliptical and lenticular, cD
\end{keywords}

Ever since the discovery of the bimodality in galaxy colour in the galaxy colour-magnitude and colour-mass diagrams from large-scale surveys \citep{2001AJ....122.1861S, 2004ApJ...600..681B, 2006MNRAS.373..469B}, the colour space between the two main populations --- the \textit{green valley} --- has been viewed as the crossroads of galaxy evolution. The galaxies in the green valley were thought to represent the transition population between the blue cloud of star-forming galaxies and the red sequence of quenched, passively evolving galaxies \citep[e.g.,][]{2004ApJ...608..752B, 2007ApJS..173..293W, 2007ApJS..173..315S, 2007ApJS..173..342M, 2007ApJ...665..265F,2011ApJ...736..110M,2012ApJ...759...67G}. Roughly speaking, all galaxies were presumed to follow similar evolutionary tracks across the green valley, with a fairly rapid transition implied by the relative scarcity of galaxies in the green valley compared to the blue cloud or red sequence.

The intermediate galaxy colours of green valley galaxies have been interpreted as evidence for the recent quenching of star formation \citep{2007ApJS..173..267S}. 
The clustering of active galactic nuclei (AGN) host galaxies in the green valley further suggested a role for AGN feedback in particular \citep[e.g.,][]{2007ApJ...660L..11N,2008A&A...490..905H, 2008ApJ...675.1025S}. Galaxies in the green valley have specific star formation rates (sSFR) lower than the ``main sequence" of star formation in galaxies, which is a tight correlation between galaxy stellar mass and star formation rate, presumably as a result of quenching \citep[e.g.][]{2004MNRAS.351.1151B, 2007A&A...468...33E, 2007ApJS..173..267S, 2007ApJ...660L..43N, 2010ApJ...721..193P, 2011A&A...533A.119E,2012ApJ...752...66L, 2012ApJ...745..149L}). Most star-forming galaxies live on the main sequence, so tracing the populations leaving the main sequence -- those with lower sSFRs -- probes the quenching mechanism(s) and, as \cite{2010ApJ...721..193P} showed, there may be at least two very different quenching processes.

Ultraviolet light comes predominantly from newly formed massive stars, which makes UV observations an excellent probe of the current rate of star formation. In this paper, we use UV-optical colours from \textit{GALEX} photometry \citep[see e.g.,][]{2007ApJS..173..342M} to investigate the rate at which galaxies are decreasing their sSFR (i.e., how rapidly they change colour), and whether this correlates with morphology.  Essentially, we use galaxy colours as stellar population clocks, an approach first conceived and applied by Tinsley and collaborators \citep[][]{1968ApJ...151..547T,1976ApJ...203...52T,1978ApJ...221..554T}. 

We interpret the evolutionary tracks of disks in terms of the gas supply and how star formation depletes the gas reservoir \citep{1959ApJ...129..243S}. Interestingly, what had appeared to be outliers from the general parent galaxy population --- namely, blue early types \citep{2009MNRAS.396..818S} and red late types \citep{2010MNRAS.405..783M} --- far from being curiosities, are instead a valuable clue to galaxy evolution.  

Morphology has not previously been a major ingredient in interpretations of the colour-mass diagram. Now that reliable morphological classifications have been made possible by citizen scientists in the Galaxy Zoo project \citep{2008MNRAS.389.1179L,2011MNRAS.410..166L},  we are able to investigate the relation of galaxy morphology to colour and mass. We also consider galaxy content and environment, now that standardized information is available for large galaxy samples \citep[e.g.][]{2004ApJ...600..681B,2007ApJ...671..153Y,2010ApJ...721..193P}.

The Galaxy Zoo data have already enabled many insights about the link between galaxy evolution and colour \citep{2009MNRAS.396..818S, 2010MNRAS.405..783M} and about the link between galaxy evolution and environment \citep[e.g.,][]{2009MNRAS.393.1324B,2009MNRAS.399..966S}, mergers \citep{2010MNRAS.401.1552D,2010MNRAS.401.1043D, 2012MNRAS.419...70K,2012ApJ...753..165T}, unusual galaxy types \citep{2009MNRAS.399.1191C, 2009MNRAS.399..129L, 2012MNRAS.420..878K}, and specific morphological features such as bars \citep[e.g.,][]{2011MNRAS.411.2026M, 2011MNRAS.415.3627H, 2012MNRAS.423.1485S, 2012MNRAS.424.2180M, 2013ApJ...779..162C, 2014MNRAS.tmp...97M}.
In this paper, Galaxy Zoo morphologies provide the key to understanding that early- and late-type galaxies,  even those with similar green optical colours, follow distinct evolutionary trajectories involving fundamentally different quenching modes.

Throughout this paper, we use a standard $\Lambda$CDM Cosmology 
($\Omega_{m} = 0.3$, $\Omega_{\Lambda} = 0.7$ and $\mathrm{H_{0}} = 70~\mathrm{km^{-1}~s^{-1}}$), 
consistent with observational measurements \citep{2011ApJS..192...18K}. All magnitudes are in the AB system.

%-----------------------------------------------------------------------------------------------------------------------------------

%-----------------------------------------------------------------------------------------------------------------------------------
\begin{figure*}
\begin{center}

\includegraphics[width=0.99\textwidth]{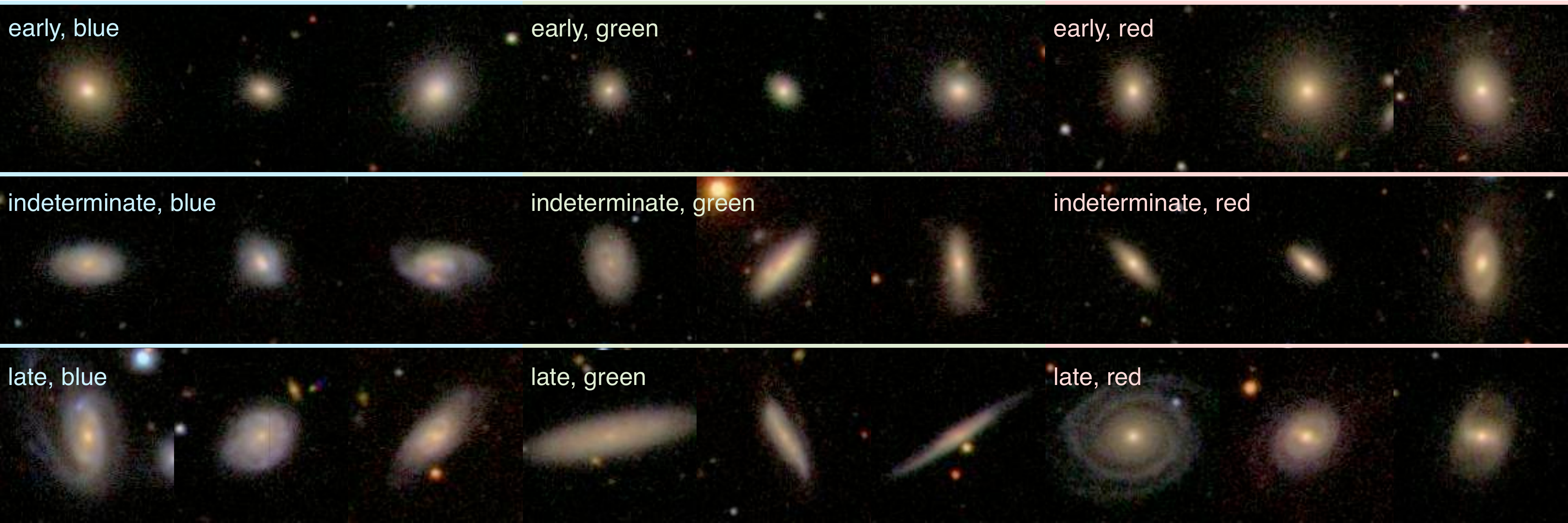}

\caption{Example $gri$ SDSS images (51.2\arcsec\ $\times$ 51.2\arcsec), ordered by Galaxy Zoo classification. In the \textit{top}, \textit{middle} and \textit{lower} rows are early-, indeterminate- and late-type galaxies, respectively. In each row, we show (from 
left to right) three blue-cloud, three green-valley and three red-sequence galaxies. The indeterminate-type galaxies are mostly composite bulge-disk systems that more closely resemble the late-type galaxies than the purely spheroidal early types. For this reason, it is not surprising that they mostly follow the late-type galaxies in their quenching behaviour.}

\label{fig:gallery}

\end{center}
\end{figure*}
%-----------------------------------------------------------------------------------------------------------------------------------

%-----------------------------------------------------------------------------------------------------------------------------------
\begin{figure*}
\begin{center}

\includegraphics[width=0.99\textwidth]{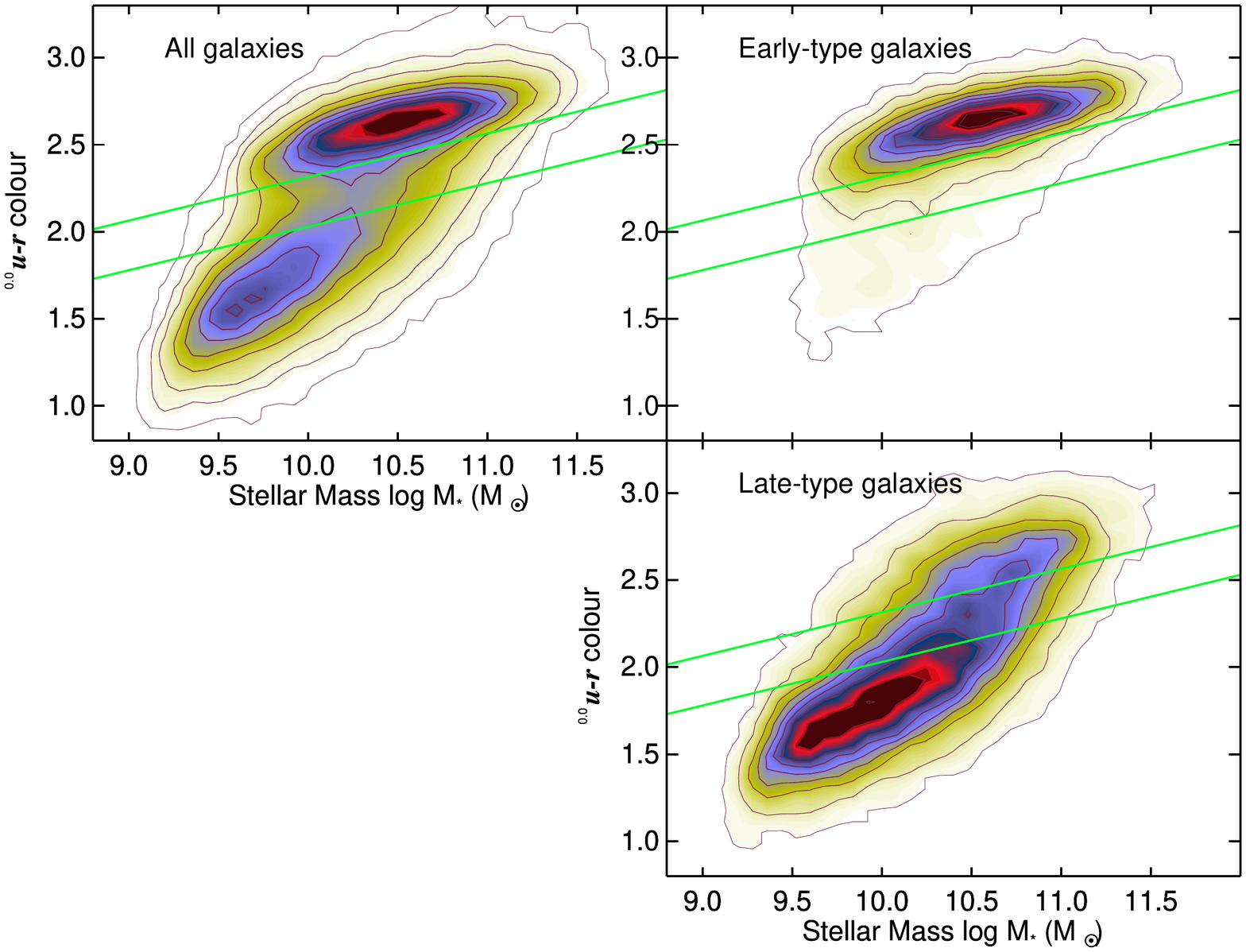}

\caption{The $u-r$ colour-mass diagram for our sample. In the {\it top left}, we show all galaxies, whereas on the {\it right}, we show the early-type ({\it top}) and late-type galaxies ({\it bottom}); green lines show the green valley defined by the all-galaxy diagram. This figure illustrates two important findings: (1) Both early- and late-type galaxies span almost the entire $u-r$ colour range. Visible in the morphology-sorted plots are small numbers of blue early-type ({\it top}) and red late-type ({\it bottom}) galaxies \citep[e.g.,][]{2009MNRAS.396..818S,2010MNRAS.405..783M}. (2) The green valley is a well-defined location only in the all-galaxies panel ({\it upper left}). Most early-type galaxies occupy the red sequence, with a long tail (10\% by number) to the blue cloud at relatively low masses; this could represent blue galaxies transiting rapidly through the green valley to the red sequence. More strikingly, the late types form a single, unimodal distribution peaking in the blue (these are the main sequence star-formers) and reaching all the way to the red sequence, at higher masses, with no sign of a 
green valley (in the sense of a colour bimodality). The contours on this Figure are linear and scaled to the highest value in each panel.
}

\label{fig:colour_mass_gv}

\end{center}
\end{figure*}
%-----------------------------------------------------------------------------------------------------------------------------------

%-----------------------------------------------------------------------------------------------------------------------------------
\section{Data}
\label{sec:data}

\subsection{Catalog generation, SDSS, and multi-wavelength data}
We briefly describe the data used in this paper. The galaxy sample is based on public photometric and spectroscopic data products from the Sloan Digital Sky Survey (SDSS) Data Release 7 \citep{2000AJ....120.1579Y, 2009ApJS..182..543A}. The initial sample selection and properties are described in \cite{2010ApJ...711..284S} and this catalogue is available on the web\footnote{See \texttt{http://data.galaxyzoo.org/}.}. The sample is limited in redshift to $0.02<z<0.05$ and limited in absolute luminosity to $M_{z,\rm{Petro}} <-19.50$ AB, 
in order to create an approximately mass-limited sample. We adopt K corrections to z=0.0 from the NYU-VAGC \citep{2005AJ....129.2562B, 2008ApJ...674.1217P, 2007AJ....133..734B}. The typical $u-r \rightarrow ~^{0.0}u-r$ K correction is $\sim0.05$ mag, and omitting the correction does not change any results significantly. We also obtain near-IR magnitudes from 2MASS \citep{2006AJ....131.1163S} via the NUY-VAGC.

We obtain aperture- and extinction-corrected star formation rates and stellar masses from the MPA-JHU catalog \citep{2003MNRAS.341...33K, 2004MNRAS.351.1151B}, which are calculated from the SDSS spectra and broad-band photometry. The spectroscopic classifications, especially the AGN-classifications, derive from analysis with the GANDALF (Gas AND Absorption Line Fitting) code \citep{2004PASP..116..138C, 2006MNRAS.366.1151S}. Ultraviolet photometry for 71\% of our sample comes from the \textit{Galaxy Evolution Explorer} (GALEX; \citealt{2005ApJ...619L...1M}), matched via the Virtual Observatory. Observed optical and ultraviolet fluxes are both corrected for dust reddening using estimates of internal extinction from the public\footnote{See \texttt{http://gem.yonsei.ac.kr/$\sim$ksoh/wordpress/}.} stellar continuum fits performed by \cite{2011ApJS..195...13O}, applying the \cite{1989ApJ...345..245C} law.

\subsection{Galaxy Zoo visual morphology classifications}
\label{ssec:gzmorph}

We use Galaxy Zoo 1 visual classifications of galaxy morphologies\footnote{Data publicly available at \texttt{http://data.galaxyzoo.org}.} from the Galaxy Zoo citizen science project \citep{2008MNRAS.389.1179L, 2011MNRAS.410..166L}. Using the \texttt{clean} criterion developed by \cite{2008MNRAS.388.1686L}, which assigns a morphology to a galaxy when 80\% or more of Galaxy Zoo users agreed on the classification, we find that for out sample, 18\% are early types, 34\% are late types and 45\% are indeterminate-types. The remaining 3\% are mergers. 

Because we restrict our analysis to galaxies with clearly determined morphologies, it is important to understand what the (large) indeterminate category represents. Either these galaxies are composite bulge-disk systems in which neither the bulge nor disk clearly dominates, or the imaging data are not good enough for a clear classification. Inspection shows the former explanation likely accounts for the vast majority of the category, 
meaning we cannot classify these systems better even with deeper imaging. Figure \ref{fig:gallery} shows example images of early-, indeterminate- and late-type galaxies. In terms of the Hubble tuning fork, the indeterminate types represent galaxies near the S0/Sa locus.  For the most part the indeterminate-morphology galaxies follow the trends of the late types, with only a small fraction being misclassified early types. We discuss this in more detail in Section \ref{sec:indet}.

Tables are cross-matched using the Virtual Observatory via TOPCAT \citep{2005ASPC..347...29T,2011ascl.soft01010T}.
%-----------------------------------------------------------------------------------------------------------------------------------

%-----------------------------------------------------------------------------------------------------------------------------------
\begin{figure*}
\begin{center}

\includegraphics[width=0.99\textwidth]{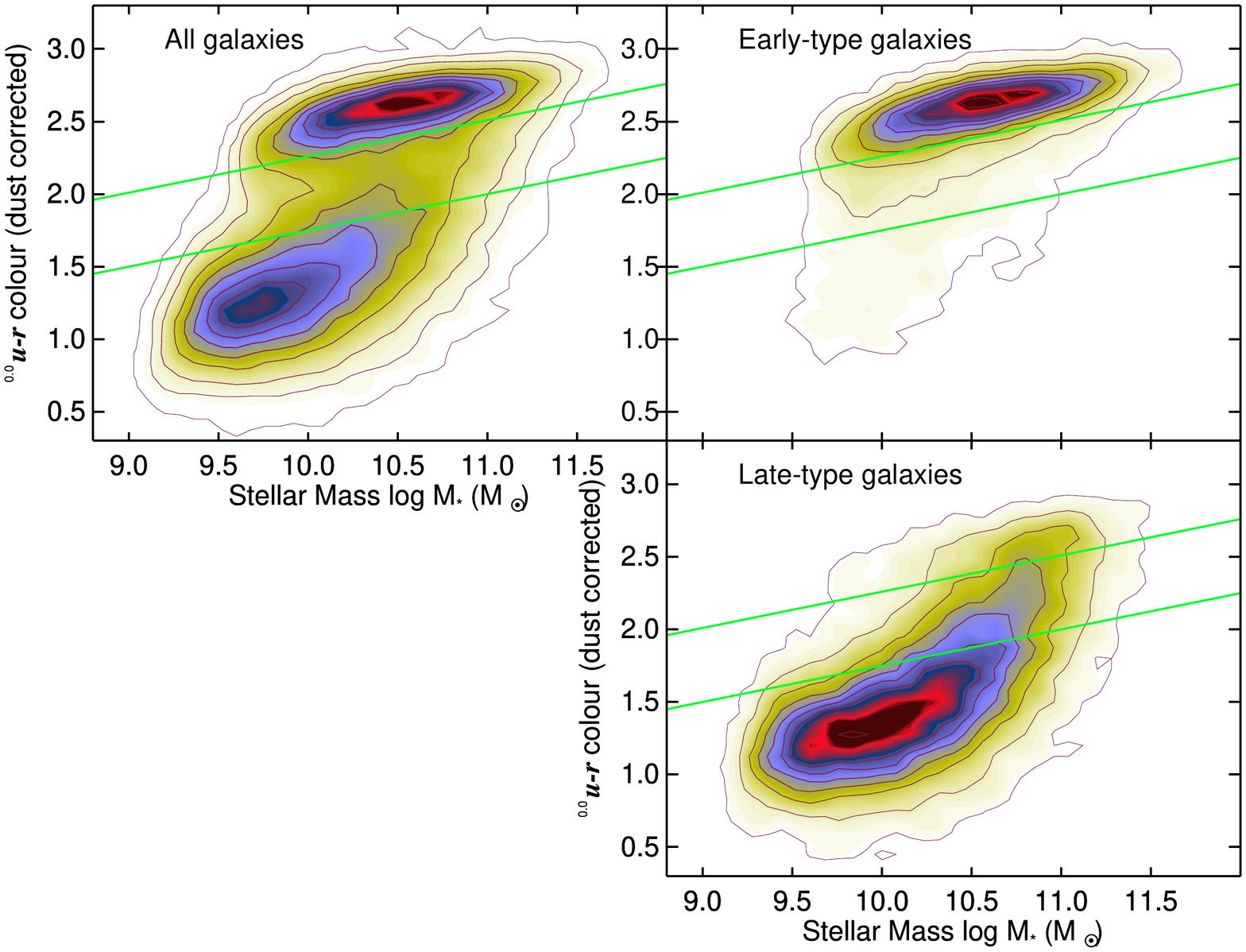}

\caption{The reddening-corrected $u-r$ colour-mass diagram for our sample. Same as Figure \ref{fig:colour_mass_gv}, but the $u-r$ colour is corrected by the E(B-V) in the stellar continuum, as measured in the SDSS spectra using the GANDALF code \citep{2011ApJS..195...13O}. 
Compared to Figure \ref{fig:colour_mass_gv}, there are no significant changes; in particular, very few green or red late-type galaxies 
actually belong in  the blue cloud. The slope of the disk galaxy contours becomes more horizontal, making clear that late types evolve more slowly than early-type galaxies. Moreover, there is clearly a tail of galaxies rising above the blue cloud at high masses, whereas the blue tail of the early types is toward low masses.  While some red late types are indeed dust-reddened, intrinsically blue galaxies, many are not, and the overall sense remains that the colours of late-type galaxies change slowly. The green valley defined here, 
from the all-galaxies panel ({\it upper left}), is used throughout the rest of this paper. The contours on this Figure are linear and scaled to the highest value in each panel.}

\label{fig:colour_mass_gv_nodust}

\end{center}
\end{figure*}
%-----------------------------------------------------------------------------------------------------------------------------------

%-----------------------------------------------------------------------------------------------------------------------------------
\section{A journey through the green valley: two evolutionary pathways for quenching star formation}
\label{sec:gv}

In this section, we look at how star formation varies in galaxies and consider variables that might affect star formation. 
We present the well-known colour-mass diagram, first as it is observed for our galaxy sample (\S~3.1.1), then after correcting for dust reddening (\S~3.1.2). In both cases, sorting by morphology dramatically changes the impression of bimodality and thus drives a new interpretation of the green valley. We then present other observables relevant to characterizing galaxy evolution: UV constraints on current star formation (\S~3.3), environment density and halo mass (\S~3.5), atomic gas reservoir (\S~3.6), and black hole growth (\S~3.7).

Most star-forming galaxies exhibit a tight, nearly linear correlation (henceforth referred to simply as the ``main sequence") between galaxy stellar mass and star formation rate, which changes with redshift only in its normalisation (at least out to $z\sim 2$, perhaps out to $z\sim4$; \citealt{2004MNRAS.351.1151B, 2007A&A...468...33E, 2007ApJS..173..267S, 2007ApJ...660L..43N, 2010ApJ...721..193P, 2011A&A...533A.119E,2012ApJ...752...66L, 2012ApJ...745..149L}). This correlation between galaxy mass and star formation rate is likely the result of an equilibrium between galaxy inflows and outflows (see \citealt{2010ApJ...718.1001B} and the ``bathtub" model of \citealt{2013arXiv1303.5059L}). Star-forming galaxies live on the main sequence regardless of whether they have spent a long time on it or have only recently re-started star formation. 
Accordingly, spending only a short time on the main sequence erases most of the past star formation history (in terms of galaxy colours). 
Then, when star formation is quenched, galaxies leave the main sequence, and we can interpret their changing colours as a reflection of the quenching process.

\subsection{Galaxy colour bimodality as a function of morphology}
\label{sec:gv_bimod}
We begin by showing that the green valley is not a single, unified population of galaxies, but rather a superposition of two populations that happen to exhibit the same intermediate (i.e., green) optical colours. The green valley is the space in the colour-mass diagram between the blue cloud and the red sequence; below we give a precise definition of the green valley in terms of  $u-r$ colour.  The interpretation of intermediate galaxy colours in terms of star formation histories is not original here; for example, it has been argued previously by \cite{2009AIPC.1201...17S}, \cite{2012arXiv1206.6496C}, and \cite{2012arXiv1206.5807C}.

\subsubsection{The colour-mass diagram}

The observed  $u-r$ colour-mass diagrams of galaxies by morphology at $z\sim0$ are shown in Figure \ref{fig:colour_mass_gv}. Contours in each panel show the linear density of galaxies and green lines indicate the location of the green valley, defined from the all-galaxy panel at the upper left. The right-hand panels show only early types (\textit{top}) or late types (\textit{bottom}). These colour-mass diagrams, which constitute one of the two main starting points of our analysis, lead us to two important findings:

\begin{enumerate}
\item \textit{Both early- and late-type galaxies span almost the entire $u-r$ colour range;} 
that is, the classification by morphology reveals populations of blue early-type galaxies and of red late-type galaxies 
\citep[e.g.,][]{2009MNRAS.396..818S,2010MNRAS.405..783M}. 
\item \textit{The green valley appears as a dip between bimodal colours only in the all-galaxies panel; within a given morphological class, there is no green valley}, just a gradual decline in number density. 
Most early types lie in the red sequence with a long tail of $\sim10\%$ of the population reaching the blue cloud, which could represent a population in rapid transition, commensurate with the original idea of the green valley as a transition zone. 
The late-type galaxies, however, do not separate into a blue cloud and a red sequence, but rather form a continuous population ranging from blue to red without a gap or valley in between.
\end{enumerate}

The traditional interpretation (and visual impression) from the all-galaxies diagram --- that blue star-forming galaxies evolve smoothly and quickly across the green valley to the red sequence --- changes when viewed as a function of morphology. 

Specifically, the impression of bimodality in the all-galaxies colour-magnitude diagram depends on the superposition of two separate populations: 
late types that are mostly in the blue cloud, decreasing smoothly all the way to the red sequence, and early types, a few of whose colours reach all the way to the blue cloud. Consideration of the indeterminate morphology galaxies (see \S~\ref{sec:indet}) actually strengthens this conclusion, as they are mostly blue disks with prominent red bulges, hence the green colours.

The blue late-type galaxies, in particular, show no signs of rapid transition to the red sequence; indeed, they must take a very long time to reach the red sequence (\S~\ref{sec:quench_model}). The early types do appear to transition quickly across the green valley, in that there are few of them with green colours and even fewer with blue colours.  This suggests the bluest early types might have been produced by major mergers of late types.

The demographics of galaxies by colour and morphology in Table \ref{tab:demo} make the point about evolutionary time scale very clearly (for the moment ignoring changes from one morphology into the other): early types spend most of their time on the red sequence, while late types remain in the blue cloud for most of their lifetimes.

%-----------------------------------------------------------------------------------------------------------------------------------
\begin{figure}
\begin{center}

\includegraphics[width=0.49\textwidth]{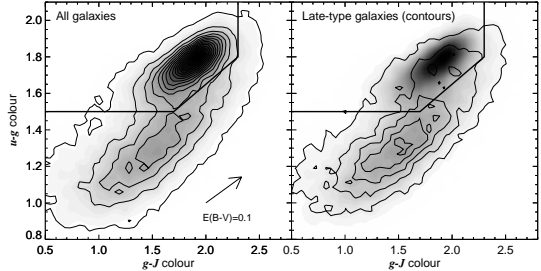}

\caption{The $ugJ$ colour-colour diagram for our sample, analogous to the $UVJ$ diagram of \citet{2009ApJ...691.1879W}. The arrow indicates the shift due to dust for an E(B-V)=0.1 using a \citet{2000ApJ...533..682C} extinction law. In the \textit{left}-hand panel, we show all galaxies and an adapted box separating passive red galaxies from dusty star-forming galaxies. The \textit{right}-hand panel shows the (same) shading of the entire galaxy sample, with contours for late-type galaxies only; this shows that some of the red spirals are actually dust-reddened spirals, while others are passively evolving.}

\label{fig:uggj}

\end{center}
\end{figure}
%-----------------------------------------------------------------------------------------------------------------------------------

%-----------------------------------------------------------------------------------------------------------------------------------
\begin{table}
\begin{center}
\caption{Demographics of galaxies in the blue cloud, green valley and red sequence by morphology}
\label{tab:demo}
\begin{tabular}{@{}lcc}
\hline
 \multicolumn{1}{l}{Galaxy} 				& \multicolumn{1}{l}{N} 		& \multicolumn{1}{l}{\%} 		\\
 \multicolumn{1}{l}{Sample} 				& \multicolumn{1}{l}{} 		& \multicolumn{1}{l}{of population} 	\\
\hline
\hline
 Early-type, blue cloud 			& 464 		& 5.2\% \\
 Early-type, green valley 		& 1,110 		& 12.4\% \\
 Early-type, red sequence 		& 7,404	 	& 82.5\% \\
\hline
 \textit{Early-type, all}				& 8,978		& 100\% \\
\\ 

 Late-type, blue cloud 			& 12,380 	& 74.1\% \\
 Late-type, green valley 			& 3,152 	& 18.9\% \\
 Late-type, red sequence 		& 1,175 	& 7.0\% \\
\hline
 \textit{Late-type, all}				& 16,707		& 100\% \\
 \hline
\end{tabular}
\\
\begin{flushleft}
%$^a$ Sample footnote.\\
\end{flushleft}
\end{center}
\end{table}
%-----------------------------------------------------------------------------------------------------------------------------------

\subsubsection{The extinction-corrected colour-mass diagram}
\label{sec:gv_dust}

Dust extinction reddens galaxies, and significant reddening from blue to red has been reported for high-redshift galaxies \citep[e.g.,][]{2009ApJ...706L.173B, 2009ApJ...691.1879W, 2010ApJ...721L..38C}, although this effect should be of limited importance at low redshift, where specific star formation rates and gas fractions are lower. Nevertheless, since we are focusing on the quenching of star formation, we must first assess the effect of dust in moving intrinsically blue galaxies from the blue cloud to the green valley or the red sequence. In particular, significant amounts of dust in inclined spirals have been reported by \cite{2010MNRAS.404..792M}, and \cite{2013arXiv1306.6552S} have shown that the reddest galaxies in the local universe are edge-on disks.

We use the $U-V$ versus $V-J$ approach introduced by \cite{2009ApJ...691.1879W} to separate dusty red galaxies from passive red galaxies. In Figure \ref{fig:uggj}, we show the very similar $u-g$ versus $g-J$ diagram (using SDSS+2MASS data), from which we conclude that there are some dusty, star-forming galaxies at low redshift (upper right side of \textit{left} panel). Most of these are late-type galaxies (\textit{right} panel) and many are highly inclined spirals (as traced by the b/a axis ratio; see also \citealt{2010MNRAS.404..792M}). \cite{2013arXiv1303.4409M} similarly found that the dusty starburst part of the $U-V$ versus $V-J$ diagram at low redshift contains some objects, but far fewer than at high redshift.

Even though the effect is small, we correct the $u-r$ colours using an estimate of the extinction in the stellar continuum. We take the measured E(B-V) values from \citealt{2011ApJS..195...13O} \citep[based on the GANDALF code][]{2004PASP..116..138C, 2006MNRAS.366.1151S}, and use the \cite{2000ApJ...533..682C} extinction law; for the \textit{GALEX} magnitudes, we use the \cite{1989ApJ...345..245C} law. We show this dust-corrected colour-mass diagram in Figure \ref{fig:colour_mass_gv_nodust}.

The main differences after  correcting for dust reddening are that the blue cloud (i.e., the main sequence) is now bluer, and the slope to redder colours with increasing mass flattens (presumably driven by dust from higher SFRs). The separation of the blue cloud and red sequence also becomes more prominent. Vitally, the green valley population in both the early- and late-type population does not disappear, and red late-type galaxies remain; not all of them were dusty starformers. Thus, dust correction is important but does not greatly change the global picture.

We now define the green valley population on the dust-corrected colour-mass diagram for all galaxies (upper left panel of Figure \ref{fig:colour_mass_gv_nodust}):

\begin{equation}
^{0.0}u-r(M_{\rm stellar})= -0.24 + 0.25\times M_{\rm stellar} ,
\end{equation}

\begin{equation}
^{0.0}u-r(M_{\rm stellar}) = -0.75 + 0.25 \times M_{\rm stellar} .
\end{equation}

\noindent
We refer to galaxies satisfying this colour criterion as \textbf{green valley} galaxies, even as we argue this term does not have a simple physical meaning. Table \ref{tab:demo} presents general demographic information about the early- and late-type galaxies. 
These results are insensitive to adjustments of the specific boundaries of the green valley.

\subsection{The different recent star formation histories of early- and late-type galaxies}

We now consider \textit{why} galaxies are in the green valley. The analysis in Section \ref{sec:gv_dust} shows that dust extinction is not the main reason. Instead, we show here that early- and late-type galaxies have very different recent star formation histories which result, coincidentally, 
in the same green valley colours.

%-----------------------------------------------------------------------------------------------------------------------------------
\begin{figure*}
\begin{center}

\includegraphics[width=0.99\textwidth]{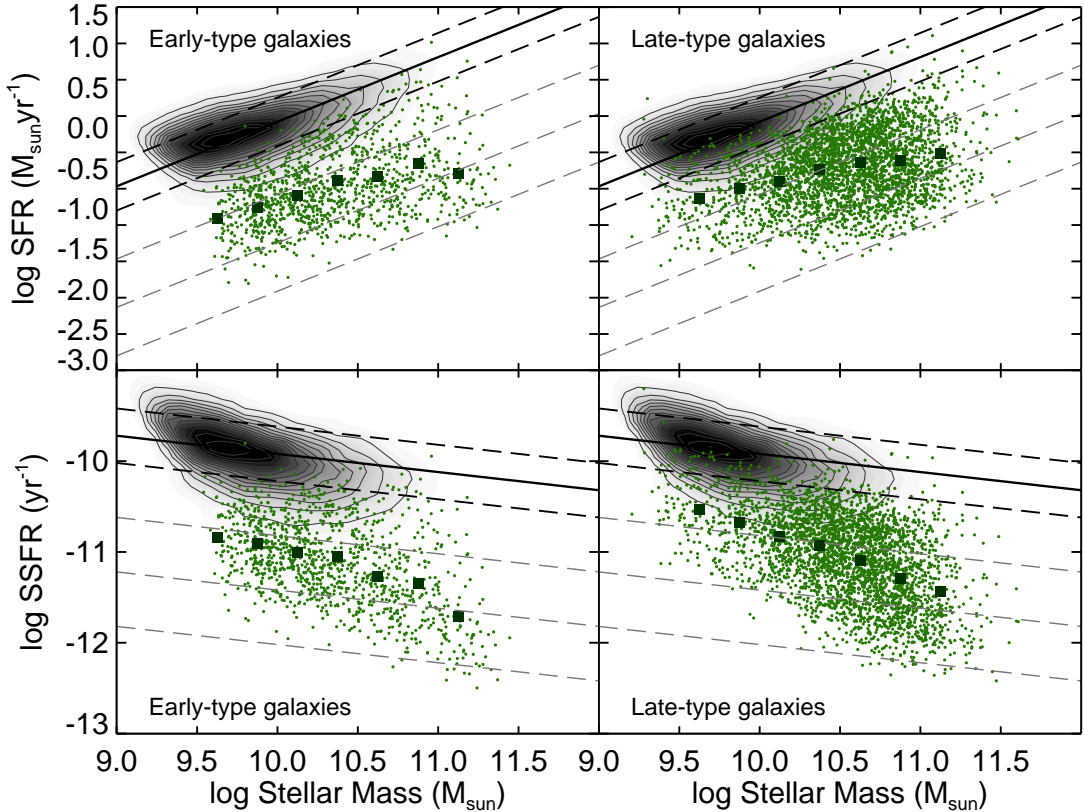}

\caption{The star formation rate (SFR) and specific star formation rate (SSFR) versus stellar mass diagrams, which highlight the main sequence. In each panel, grey-shaded contours show galaxies classified as star-forming according to the BPT emission line diagram 
(regardless of morphology); lines indicate the main sequence ({\it solid}) and 0.3 dex scatter ({\it dashed}) with $\beta=-0.2$ (the DR7 $\beta$ is lower than the DR4 value of $-0.1$; Y. Peng, priv. comm.); the grey dashed lines show further 0.3 dex offsets down from the main sequence. 
The \textit{top} row shows the standard star formation rate versus stellar mass diagram and the \textit{bottom} row shows the specific star formation rate instead. In the \textit{left}-hand column, we show the green valley early types as green points and in the \textit{right}-hand column, we show the green valley late types as green points. The large green squares are median values. For both green valley populations, we plot SFR/sSFRs as reported by the MPA-JHU catalogue regardless of whether the object is classified as star-forming or not
(i.e., including upper limits). Both populations are clearly offset from the general population of main sequence star-forming galaxies --- as expected, since they are quenching ---  and the early types tend to lie further off the main sequence than the late types, especially in the sSFR plot;
however, the difference in {\it optical} colour is small and only UV colours indicate \textit{how fast} galaxies are moving off the main sequence.}

\label{fig:ms}

\end{center}
\end{figure*}
%-----------------------------------------------------------------------------------------------------------------------------------

%-----------------------------------------------------------------------------------------------------------------------------------
\begin{figure*}
\begin{center}

\includegraphics[width=0.99\textwidth]{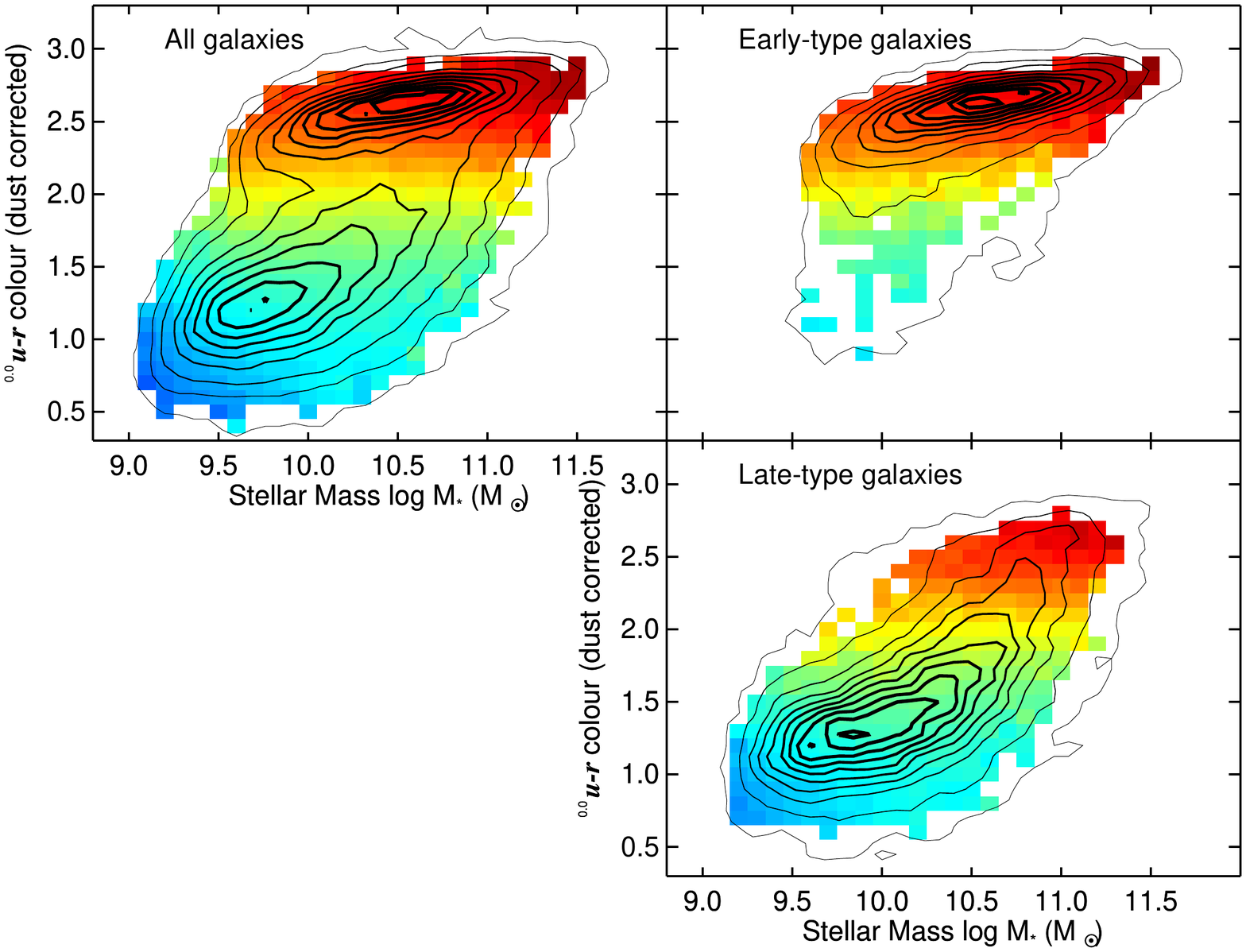}

\caption{The dust-corrected colour-mass diagram, 
like
Figure \ref{fig:colour_mass_gv_nodust} but with the galaxy populations in 0.1$\times$0.1 dex panels coloured by the mean specific star formation rate. This diagram shows that both green valley early- and late-type galaxies have lower sSFRs than their blue cloud counterparts, i.e., they are off the main sequence. Like the SFR/sSFR diagrams in Figure \ref{fig:ms}, this Figure shows that the (dust-corrected) green valley is populated by off-main sequence galaxies but \textit{it does not show} \textit{how rapidly} the sSFRs are declining. }

\label{fig:colour_mass_ms}

\end{center}
\end{figure*}
%-----------------------------------------------------------------------------------------------------------------------------------

\subsubsection{Green valley galaxies are offset from the main sequence of star formation}
\label{sec:main_seq}

Figure \ref{fig:ms} shows the stellar mass versus SFR (star formation rate -- from aperture-corrected H$\alpha$) and sSFR (specific star formation rate) diagrams, with the green valley early- and late-type galaxies highlighted. The grey contours show the main star-forming population (of all morphologies), identified spectroscopically, while the green valley objects are plotted regardless of emission line class (This means that some fraction of the SFRs and sSFRs are upper limits.)

Figure \ref{fig:ms} shows that green valley galaxies (green points) are objects that have moved off the main sequence. That is, since virtually all star-forming galaxies are on the main sequence and since green valley galaxies must have experienced star formation in the past, something has moved them off the main sequence.The solid black  line shows the local main sequence with a slope\footnote{Where $\beta$ is the exponent in sSFR $\propto$ $M_{\rm stellar}^{\beta}$.} $\beta$ of $-0.2$ (based on DR7 data; $\beta =-0.1$ for DR4; Y. Peng, priv. com), with dashed lines indicating $\pm0.3$ dex. The gray dashed lines show further 0.3 dex offsets down from the main sequence. From both the SFR and sSFR diagrams, it is apparent that both early- and late-type galaxies in the green valley are also off the main sequence --- as expected because they are in the process of quenching. What this diagram does not reveal is \textit{how fast} galaxies are moving off the main sequence.

To complete the circle, we return to the colour-mass diagram and combine it with sSFR information. In Figure \ref{fig:colour_mass_ms}, we show the (dust-corrected) colour-mass diagram, analogous to Figure \ref{fig:colour_mass_gv_nodust} except that we colour 0.1$\times$0.1 dex panels by the average sSFR in each bin. Not too surprisingly, this reveals a good correlation between dust-corrected $u-r$ colour and sSFR, showing that the green valley is, as expected, the region in colour-mass space where sSFRs have declined as galaxies have moved off the main sequence. Still, like the original colour-mass diagram, this figure does not reveal the time scales on which the sSFRs decline, so in a sense it obscures the fact (presented below) that early- and late-type galaxies transition very differently through the green valley. 

We note that the increasing prominence of bulges in massive, red late types (e.g., \citealt{2010MNRAS.405..783M}) does not significantly alter the $u-r$ colour for the present sample:  Figure \ref{fig:ms} makes it clear that all intrinsically green galaxies (the green points) are off the main sequence 
regardless of morphology. Star-forming late types on the main sequence that would appear  green in $u-r$ colour due to a luminous, red bulge
have either been excluded (such objects are likely to be classified as``indeterminate'' since they have both bulge and disk) or, if classified as bona fide late types, the young blue stars simply outshine the red bulge.

%-----------------------------------------------------------------------------------------------------------------------------------
\begin{figure*}
\begin{center}

\includegraphics[width=0.99\textwidth]{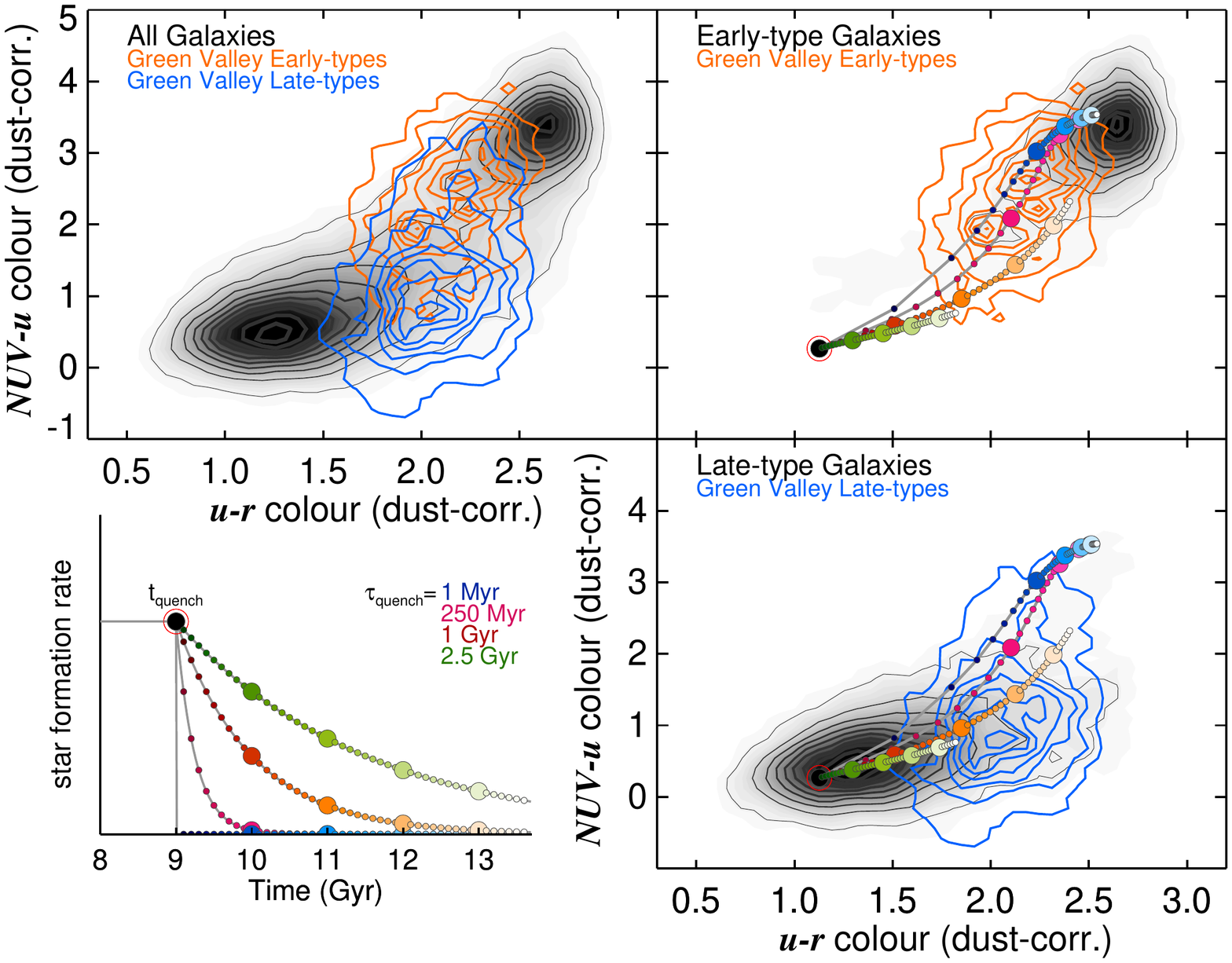}

\caption{UV-optical colour-colour diagrams (corrected for dust) 
used 
to diagnose the recent star formation histories of galaxies. Unlike the sSFR diagrams, these colour-colour diagrams constrain the rate of change in the sSFR, i.e., how rapidly star formation quenches in these galaxies. In each panel, the grey contours represent the underlying galaxy population, while the coloured contours represent galaxies 
with 
(optical) green valley
colours. 
In the {top-left} panel, we show the entire galaxy population and the early- and late-type galaxies in the green valley (orange and blue, respectively). In the {right} hand panels, we show only early-type galaxies (\textit{top}) and only late-type galaxies (\textit{bottom}). Note that early-type galaxies in the (optical) green valley are significantly redder in $NUV-u$ than late types with the same green valley (optical) colours, indicating they harbor far fewer very young stars. On top of the \textit{right}-hand panels, we plot a series of evolutionary tracks. Each track follows the same star formation history: constant star formation rate until, at a time $t_{\rm quench} =9$ Gyr, star formation begins to decline exponentially with a quenching time scale $\tau_{\rm quench}$. The lower left plot shows four such star formation histories, with an effectively instantaneous $\tau_{\rm quench}$ of 1 Megayear ({\it blue}); more moderate time scales of 250 Myrs ({\it red}) and 1 Gyr ({\it orange}); and a gentle decline with $\tau_{\rm quench} =2.5$~Gyr ({\it green}). We overplot these colour-coded evolutionary tracks on the colour-colour diagrams on the \textit{right}. For each track, we show 100 Myr intervals as small points and 1 Gyr intervals as large points to give a sense of how rapidly galaxies transit the colour-colour diagrams. These diagrams show clearly that the quenching time scales of early-type galaxies must be very rapid ($\tau_{\rm quench} \lesssim 250$~Myr), while late-type galaxies must quench very slowly ($\tau_{\rm quench} > 1$~Gyr).}

\label{fig:evol1}

\end{center}
\end{figure*}
%-----------------------------------------------------------------------------------------------------------------------------------

\subsection{UV-optical colour-colour diagrams constrain the star formation quenching time scale}
\label{sec:quench_model}
O- and B- and A-stars have very different colours and lifetimes and thus can provide leverage over the very recent star formation histories of galaxies. The SFR and sSFR diagrams, with SFRs based on H$\alpha$ line emission, provide a constraint on recent star formation properties but not on \textit{how rapidly} the (s)SFR is changing. H$\alpha$ traces the OB stars on timescales of $10^{6}-10^{7}$ years, the restframe UV traces the range of $10^{7}-10^{8}$ years while optical colours are sensitive up to $10^{9}$ years. We use a UV-optical colour-colour diagram, which takes into account the the age differential probed by the UV-optical of the SED. This UV-optical colour-colour diagram thus is sensitive to the time derivative of the SFR, to argue that the current star formation histories of green valley early- and late-type galaxies are, in fact, very different. 

Figure \ref{fig:evol1} shows the (dust-corrected) $NUV-u$ versus $u-r$ colour-colour diagrams of local galaxies. In the \textit{top-left} panel, we show the entire galaxy population (grey contours) and the green valley early- and late-type galaxies (orange and blue contours, respectively). In the \textit{top-right} panel, we show only the early-type galaxies, and in the \textit{bottom-right} panel, only the late-type galaxies, with the green valley populations again as orange and blue contours, respectively. 

Most noteworthy in Figure \ref{fig:evol1} is that, while the early- and late-type galaxies in the green valley exhibit (by selection) similar $u-r$ colours, they have significantly different $NUV-u$ colours. The early-type galaxies exhibit much redder $NUV-u$ colours \textit{at the same optical colour} than the late types in the (optical) green valley. 

This analysis shows that early-type galaxies in the (optical) green valley are quenched rapidly: they show little ongoing star formation while still having significant intermediate-age stellar populations. They feature classic post-starburst stellar populations. The late-type galaxies in the (optical) green valley, on the other hand, show similar $NUV-u$ colours as their star-forming counterparts in the ($u-r$) blue cloud. This is consistent with slowly declining star formation, so that late types have enough ongoing star formation to still be blue in the ultraviolet, yet the overall stellar population is aging (the mean stellar age is increasing), thus moving them into the optical green valley (and off the main sequence).
Indeed, the lack of a green valley in the late-type plot is further evidence for their gradual quenching.

These $NUVur$ colour-colour diagrams are clearly sensitive diagnostics of young and intermediate age stellar populations, and therefore of recent star formation histories. Using model star formation histories, we can quantify this interpretation and in particular, constrain the \textit{time scales} on which star formation declines in the two populations. We construct an illustrative star formation history as follows: a constant star formation rate for 9 Gyr followed by a transition to an exponentially declining star formation rate with variable time scale, $\tau_{\rm quench}$, representing the quenching time scale. We note that a constant star formation rate is a reasonable model for a galaxy on the main sequence: despite the fact that the sSFR drops by a factor of $\sim20$ from $z\sim1$ to today, the SFR only changes by about a factor 3 \citep{2013arXiv1303.5059L}. We discuss the robustness of this model further in Appendix \ref{appendix_a}.

We generate model star formation histories and convolve them with \cite{2003MNRAS.344.1000B} population synthesis model spectra to generate a model SED. We blank out the youngest 3 Myr stellar populations to mimic the effect of birth clouds (which obscure the youngest stellar populations), and finally, 
convolve with filter transmission curves to generate observed colours. We vary $\tau_{\rm quench}$ from 1 Myr (effectively instantaneous suppression of star formation) to 2.5 Gyr (a slow decline corresponding a quenching process significantly slower than the dynamical time scale of a galaxy). The {lower-left} panel of Figure \ref{fig:evol1} shows a schematic of these model star formation histories and the corresponding tracks are over-plotted on the $NUVur$ colour-colour diagrams on the {right}-hand panels. For each track, we mark 100 Myr intervals with a small point and 1 Gyr intervals with a large point. The $\tau_{\rm quench}=1$ Myr track moves rapidly across the diagram within $\sim1$ Gyr, while the $\tau_{\rm quench}=2.5$ Gyr track barely moves at all in several Gyr; in fact, it never leaves the blue cloud, a point to which we return later.

\begin{figure}
\begin{center}

\includegraphics[width=0.49\textwidth]{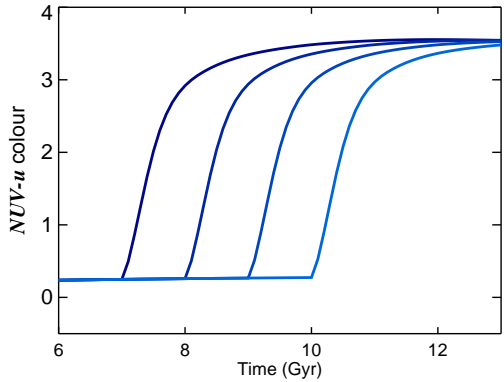}

\caption{The evolution of the $NUV-u$ colour based on BC03 models following the simple quenching model described in Section \ref{sec:quench_model}. At some time $t_{\rm quench}$, a constant SFR star formation history is interrupted and enters an exponential decline with $\tau = 100$ Myr. We vary $t_{\rm quench}$ from 7 to 10 Gyr age to show that the precise quenching time has no significant effect on the $NUV-u$ colour after the rapid  ($\sim100$~Myr)colour transition. This figure illustrates the sensitivity of the $NUV-u$ colour to short timescales in the 10-1,500 Myr range. With a dynamic range of nearly 4 mag, this colour is ideal for tracing quenching time scales.}

\label{fig:nuvu_time}

\end{center}
\end{figure}

In Figure \ref{fig:nuvu_time}, we show using \cite{2003MNRAS.344.1000B} models that the value of the time of the quenching event does not matter significantly as the $NUV-u$ colour is strongly dominated by young- and intermediate-age stellar populations in the 10-1,500 Myr time range. 

We compare these tracks to the observed locations of early- and late-type galaxies on the $NUVur$ diagram. The early-type galaxies in the green valley are well matched by tracks with very short $\tau_{\rm quench}$. Both the instantaneous truncation track and the track with $\tau_{\rm quench}=250$ Myr pass straight through the early-type green valley locus before reaching the red sequence. The $\tau_{\rm quench}=1$ Gyr track misses the main early-type locus and in fact stalls in the green valley. If early-type galaxies were quenching as slowly as $\tau_{\rm quench}=1$ Gyr, they would build up in the green valley, which is not observed (Figure \ref{fig:colour_mass_gv_nodust}).

The late-type galaxies in the green valley are inconsistent with most quenching tracks. The $\tau_{\rm quench}=1$ Gyr barely goes through the late-type blue cloud locus. Only the $\tau_{\rm quench}=2.5$ Gyr track passes through the bulk of the population. This indicates that long quenching time scales (several Gyrs) are required to explain the colour-colour evolution of late types.

We conclude from the $NUVur$ colour-colour diagram that early- and late-type galaxies follow 
very different evolutionary pathways to and through the green valley. Early-type galaxies undergo a rapid end to star formation, transiting the green valley rapidly, perhaps as rapidly as stellar evolution allows. By the time they appear in the green valley, most if not all star formation has ceased. Late-type galaxies, in contrast, experience at most a slow decline in star formation and gradual departure from the main sequence. The slowly declining star formation leads to increasingly red optical colours but not necessarily redder UV-optical colours, because extremely young, luminous stars continue to form.

%-----------------------------------------------------------------------------------------------------------------------------------
%-----------------------------------------------------------------------------------------------------------------------------------

%-----------------------------------------------------------------------------------------------------------------------------------
\begin{figure}
\begin{center}

\includegraphics[width=0.49\textwidth]{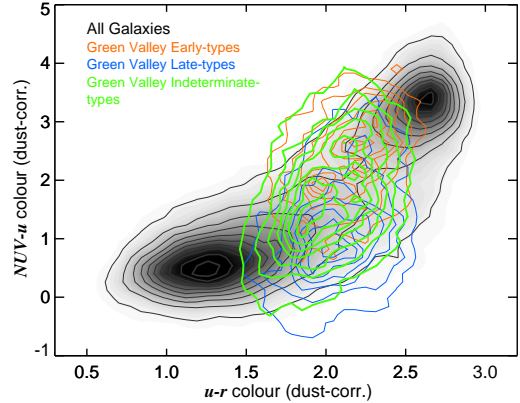}

\caption{$NUVur$ diagram, similar to the upper left panel in Figure \ref{fig:evol1} but showing the indeterminate-type galaxies as green contours, with early types as orange contours and late types as blue contours. The bulk of the green valley indeterminate-types overlap the green valley late-type locus, with somewhat redder colours, with a minority scattering to the early-type locus. This supports a picture where the indeterminate-types quenching slowly, similar to the late types, with the minority misclassified early types possibly quenching more rapidly. }

\label{fig:evol2_indet}

\end{center}
\end{figure}
%-----------------------------------------------------------------------------------------------------------------------------------

\subsection{Caveat: Indeterminate-type galaxies}
\label{sec:indet}

We now revisit the question of the large population of indeterminate-type galaxies (the galaxies that did not receive at least 80\% agreement in any morphological category). As is apparent in Figure \ref{fig:gallery}, most of the indeterminate-types show disk features. In Figure \ref{fig:evol2_indet}, we show the $NUVur$ diagram (dust-corrected) analogous to Figure \ref{fig:evol1}, but with the green valley indeterminate-type galaxies overplotted as green contours. The bulk of the green valley indeterminate-types overlap the green valley late-type locus, with somewhat redder colours, with a minority scattering to the early-type locus, no doubt because some of them have big red bulges. The indeterminate-types thus mostly quench slowly, similar to the late types, with a minority being misclassified early types, which quench rapidly.

The indeterminate-types do not appear to represent an intermediate quenching pathway between the extremes of early- and late-type galaxies. Instead, most follow the late types (likely related to also having a disk), with a minority following the early types, presumably because they are misclassified early types. A more precise investigation of how the indeterminates fit into the general picture presented here will have to rely on future, better imaging data.

%-----------------------------------------------------------------------------------------------------------------------------------
\begin{figure*}
\begin{center}

\includegraphics[width=0.99\textwidth]{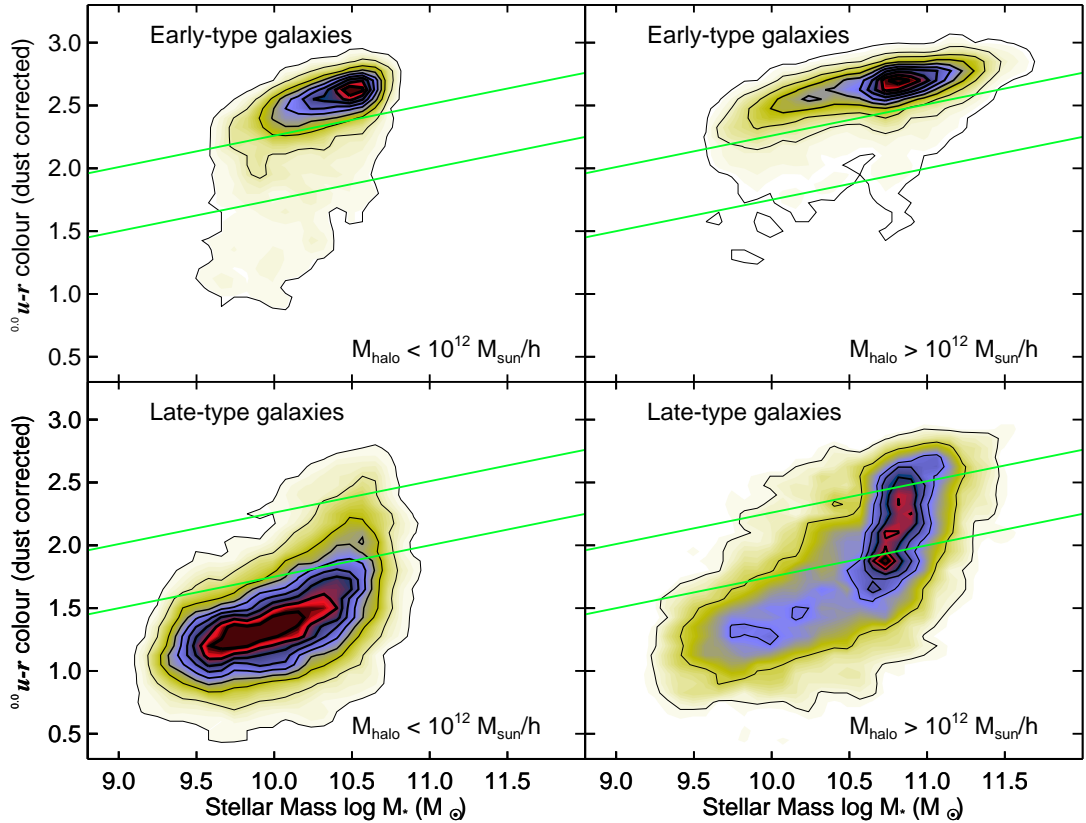}

\caption{Dust-corrected colour-stellar mass diagrams of early- and late-type galaxies (\textit{top-} and \textit{bottom} rows) split by halo mass (from \citealt{2007ApJ...671..153Y}) into low mass (\textit{left}) and high mass haloes (\textit{right}; split at $M_{\rm halo} = 10^{12} M_{\sun}/h$). Only small, qualitative differences are seen in the early-type galaxies, whereas quite striking differences appear in the late types. The green valley early types are present in both low and high-mass haloes; in low-mass haloes, they are mostly centrals, and in high mass haloes, mostly satellites. In contrast, the blue cloud late types are mostly in low-mass haloes, while the green valley late types are mostly in massive haloes, and are largely centrals.This makes clear that quenching of star formation in late-type galaxies is closely related to halo mass; those with $M_{\rm halo} = 10^{12} M_{\sun}/h$ are partly or mostly quenched, suggesting that accretion of mass through the halo slows or stops above this value.
}

\label{fig:gv_env}

\end{center}
\end{figure*}
%-----------------------------------------------------------------------------------------------------------------------------------

%-----------------------------------------------------------------------------------------------------------------------------------
\begin{figure*}
\begin{center}

\includegraphics[width=0.99\textwidth]{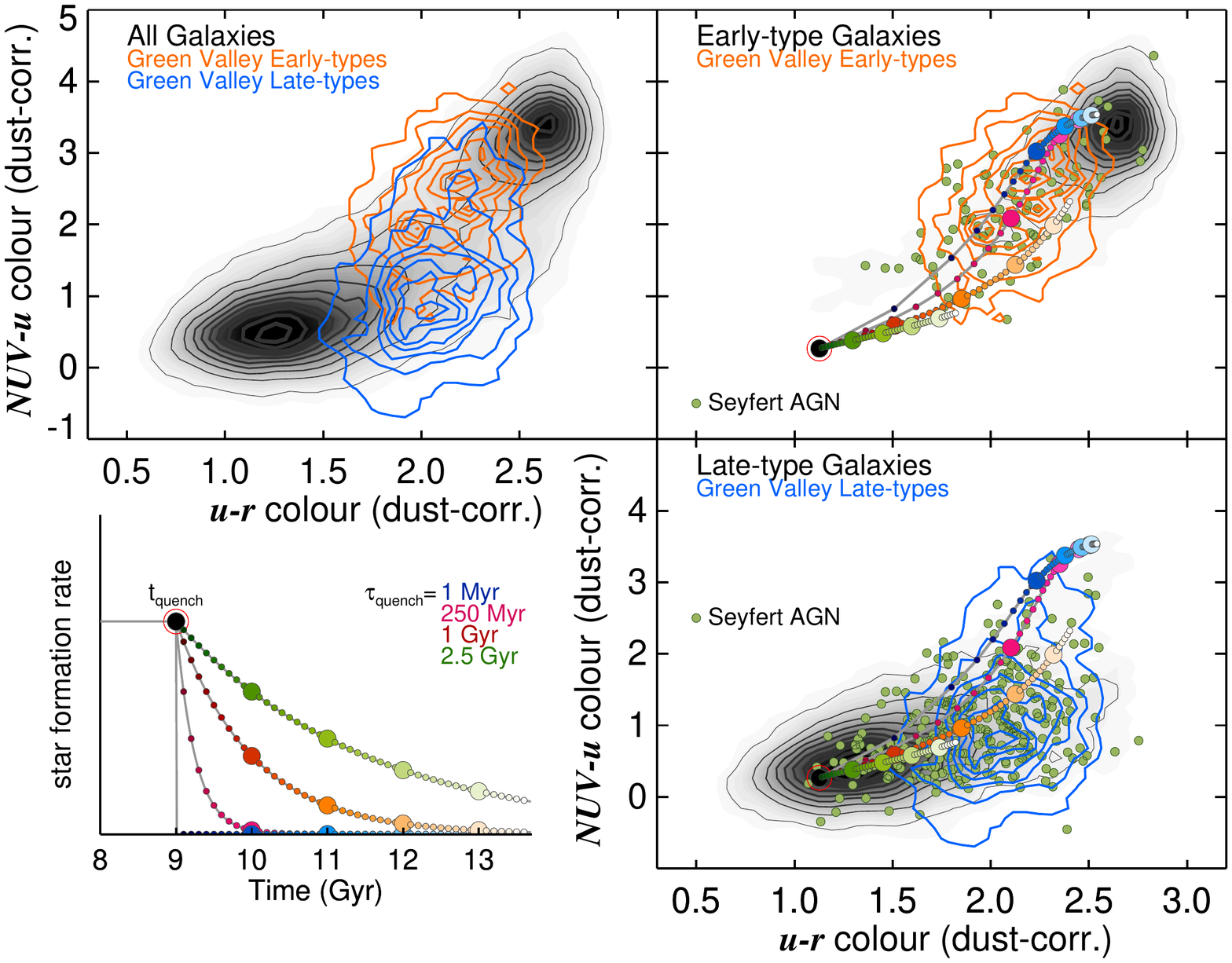}

\caption{$NUVur$ diagram, same as Figure \ref{fig:evol1}, but showing the emission line-selected AGN host galaxies as green points. This diagram places these AGN host galaxies in the context of quenching scenarios: both early-type and late-type AGN hosts lie squarely in the (optical) green valley, a few hundred Megayears or longer after the quenching event. In both cases, the AGN identified via emission lines cannot be responsible for the quenching, as they appear several hundred Megayears along the quenching tracks. }

\label{fig:evol2_agn}

\end{center}
\end{figure*}
%-----------------------------------------------------------------------------------------------------------------------------------

\subsection{Local environment, halo mass and satellite fraction of galaxies in the green valley}
\label{sec:env_gv}

We now investigate whether the environments of early- and late-type galaxies in the green valley can be linked to their very different recent star formation histories. We use the \cite{2007ApJ...671..153Y} group catalogue which yields a statistical estimate of the halo mass for each galaxy group, and whether any galaxy is the most massive/luminous in the group (central vs. satellite).

In Figure \ref{fig:gv_env}, we show the colour-mass diagram of both early- (\textit{top}) and late-type ({\textit{bottom}) galaxies split by halo mass (at $M_{\rm halo} = 10^{12}$\Msun/h). The value $10^{12}$~\Msun is motivated by previous work on halo quenching \citep[e.g.,][]{2006MNRAS.370.1651C, 2006MNRAS.368....2D}.

We find a striking difference between the early- and late-type galaxies. The green valley early-type galaxies are present in both low- and high-mass haloes. The late types show a very dramatic split: the blue cloud (i.e., main sequence) late types are mostly in low-mass haloes, while the green valley and red sequence late types (i.e., partially or mostly quenched) are almost exclusively in high mass haloes. In other words, early types 
quench in all environments whereas the quenching in late types is clearly different above and below a
halo mass of $10^{12}$~\Msun/h. 

Similar results were reported previously by \cite{2009MNRAS.399..966S}, who found that late-type quenching is associated with environment, and that late types may be able to quench without an associated morphological transformation. For an in-depth discussion of the stellar mass to halo mass relationship, see \cite{2010ApJ...717..379B}

\subsection{Atomic hydrogen gas in green valley galaxies}
\label{sec:HI_gas}

We now turn to another aspect of quenching: the gas supply for star formation. Based on the previous sections, we would expect late-type galaxies in the green valley to retain sizable reservoirs of gas to sustain ongoing, though slowly declining, star formation, while early-type galaxies should be gas poor to account for a rapid drop in new star formation. 

In order to test this hypothesis, we matched our sample to the HI database from the 40\% Arecibo Legacy Fast ALFA Survey \citep[ALFALFA;][]{2011AJ....142..170H}. In Table~\ref{tab:hi_props}, we report the numbers of early- and late-type galaxies in the green valley 
that were covered and detected in the ALFALFA survey.

We find that 48\% of all green valley late-type galaxies were detected in HI by the ALFALFA survey, consistent with many of them retaining significant gas reservoirs. In contrast, only 8\% of green valley early-type galaxies were detected in HI, supporting the picture that their star formation was quenched rapidly by removing (or ionizing) the available gas.

On a related note, we have an ongoing program to observe the HI kinematics of green valley early-type galaxies and find that most of them show highly disturbed gas kinematics consistent with having experienced recent mergers (Wong et al. in prep).

%-----------------------------------------------------------------------------------------------------------------------------------
\begin{table}
\begin{center}
\caption{HI properties of galaxies in the green valley (from 40\% ALFALFA data)}
\label{tab:hi_props}
\begin{tabular}{@{}lcccc}
\hline
 \multicolumn{1}{l}{Sample} & \multicolumn{1}{l}{N} & \multicolumn{1}{l}{N} & \multicolumn{1}{l}{N} & \multicolumn{1}{l}{HI Detection} \\
 \multicolumn{1}{l}{Green Valley...} & \multicolumn{1}{l}{covered} & \multicolumn{1}{l}{detected} & \multicolumn{1}{l}{un-detected} & \multicolumn{1}{l}{fraction}\\
\hline
\hline
 Early-type Galaxies & 349 & 28 & 321 & 8\% \\
 Late-type Galaxies & 912 & 435 & 477 & 48\%\\
 \hline
\end{tabular}
\\
\begin{flushleft}
%$^a$ Sample footnote.\\
\end{flushleft}
\end{center}
\end{table}

%-----------------------------------------------------------------------------------------------------------------------------------

\subsection{Black hole growth and galaxies in the green valley}
\label{sec:sfh_gv}

We now turn to the question of how the growth of supermassive black holes in the centers of galaxies might be related to the separate evolutionary pathways for quenching star formation in early- and late-type galaxies. In fact, the present work provides context for interpreting our earlier Galaxy Zoo AGN host galaxy study \citep{2010ApJ...711..284S}. 

In Figure \ref{fig:evol2_agn}, we show the same $NUVur$ colour-colour diagram as in Figure \ref{fig:evol1}, with emission line-selected AGN host galaxies added as green points (see \citealt{2010ApJ...711..284S} for AGN selection details). All AGN are narrow-line Type 2 (obscured) AGN, so there should be no contribution of AGN continuum to the UV/optical colours. Both early- (\textit{top right}) and late-type (\textit{bottom right}) AGN host galaxies cluster in the (optical) green valley. This is why it has been suggested previously that black hole accretion is correlated with a decline in sSFR \citep[and therefore green colours; e.g.,][]{2007MNRAS.382.1415S, 2007MNRAS.381..543W,2007ApJ...660L..11N, 2008ApJ...675.1025S}.

Comparing to the evolutionary tracks in Figures \ref{fig:evol1} and \ref{fig:evol2_agn}, it is clear that in most cases, {\it AGN signatures become visible well past quenching time}. We make no assumption on the AGN life time here, but consider when, during the evolutionary stage of the host galaxy, black hole accretion is favoured. This delay between the shutdown of star formation and the detection of (emission-line selected) AGN has been noted previously \citep{2007MNRAS.382.1415S, 2007MNRAS.381..543W, 2007ApJ...671.1388D, 2009ApJ...692L..19S, 2009ApJ...692L..19S};  several hundred Myr or more must elapse between the end of star formation in early types and the detection of  an optical AGN (Gyrs for the late types), even assuming instantaneous quenching \citep[see discussion in][]{2007MNRAS.382.1415S, 2009ApJ...692L..19S}. This implies that the AGN radiation from green valley early-type galaxies is not responsible for the rapid quenching of star formation seen in them. Rather, the AGN 
activity is plausibly an after-effect of the event that triggered quenching, rather than its cause.

In late-type galaxies (Fig.~\ref{fig:evol2_agn}, \textit{bottom right}), black hole accretion is visible in the green valley and continues as the galaxy slowly ages to redder colours. The $NUV-u$ colour shows that many late-type galaxies --- including the AGN host galaxies --- still have young stars \citep[see also][]{2012A&A...543A.132C}. 

These statements are inferred from a sample of emission line-selected AGN; it is important to check whether AGN samples based on other (more inclusive) selections show the same trends.

%-----------------------------------------------------------------------------------------------------------------------------------

%-----------------------------------------------------------------------------------------------------------------------------------
\begin{figure*}
\begin{center}

\includegraphics[width=0.99\textwidth]{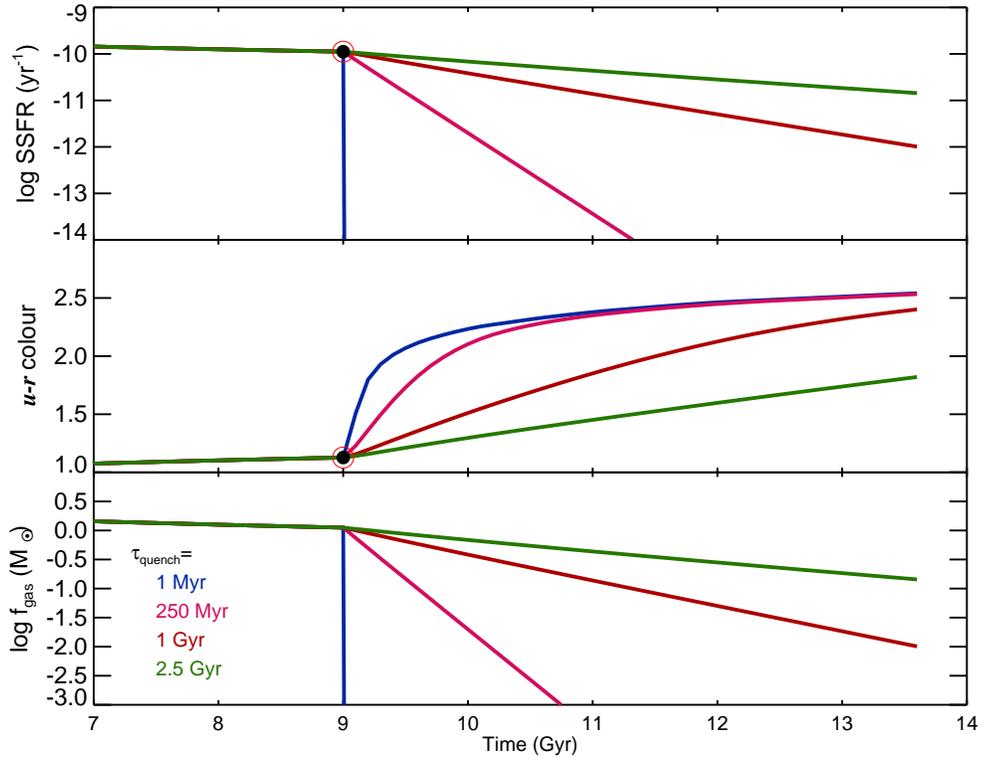}

\caption{Diagram showing the evolution of a galaxy corresponding to the quenching scenarios discussed in the text. The star formation rate is constant until quenching begins at $t=9$ Gyr, then declines exponentially with $\tau_{\rm quench}\in\{1,250,1000,2500\}$ Myr. ({\it Top panel:}) evolution of the specific star formation rate (sSFR); ({\it middle panel:}) $u-r$ colour evolution; ({\it bottom panel:}) gas fraction, inferred from the star formation rate by inverting the Schmidt law (Eqn.~\ref{eq:schmidt}). }

\label{fig:tracks}

\end{center}
\end{figure*}
%-----------------------------------------------------------------------------------------------------------------------------------

%-----------------------------------------------------------------------------------------------------------------------------------
\section{discussion: evolutionary tracks related to the end of star formation}
\label{sec:discussion}

We have used a series of observational results, informed by morphological classifications, to develop a broad picture of how and why early- and late-type galaxies at $z\sim0$ transform from blue star-forming galaxies into passively evolving red galaxies. Here we review the evidence presented in Section \ref{sec:gv}, then discuss our interpretation.

We first verified that dust extinction, while present, is not the main reason most galaxies have green colours. We then assessed the recent star formation properties of green valley galaxies, as traced by emission lines and by the location of green valley galaxies compared to the main sequence. Both indicators show that low star formation rates are the reason green valley galaxies --- regardless of morphology --- exhibit green optical colours. 

We then considered the UV-optical colours of galaxies in the optical ($u-r$) green valley, the UV (from \textit{GALEX}) being more sensitive to the youngest stars than purely optical colours. We found that, when early-type galaxies are in the (optical) green valley, their UV-optical colours are much redder than those of late-type green valley galaxies. This supports the idea that the two morphological classes have fundamentally different recent star formation histories (\textit{i.e.} different UV-u colors), even after the initial stellar populations have aged in a similar way (\textit{i.e.} optical colors are green). We used model evolutionary tracks to show that this difference corresponds to very different time scales on which star formation declines: in early-type galaxies, the quenching time scale is much shorter ($<250$~Myr) than in late-type galaxies ($>1$~Gyr).

All this evidence leads to a coherent physical picture where a quenching event destroys the (quasi) equilibrium state of main sequence star-forming galaxies, moving them off the main sequence into the green valley. The data clearly show that {\it the time scale for this colour evolution is strongly tied to morphology}. We now argue that two quenching pathways --- for early-type and for late-type galaxies --- can be tied to two different scenarios for destroying gas reservoirs, i.e., to two different physical quenching mechanisms. Because most star-forming galaxies start on the main sequence of star formation, where a simple regulator balances cosmological inflows of gas with outflows \citep{2013arXiv1303.5059L}, we take the main sequence as the starting point of the evolutionary models discussed below.

\subsection{The evolution of late-type galaxies} 
 
We have shown that the star formation histories of late-type galaxies are consistent with a very gradual quenching, 
corresponding to an exponential time scale of a gigayear or more. A natural explanation is that the baryon supply is disrupted, so that the star formation rate declines slowly in response to the depletion of the gas reservoir. The disruption could be due to halo mass quenching \citep[e.g.,][]{2006MNRAS.370.1651C}, the end of cold streams and the development of a hot halo due to shocks \citep[]{2006MNRAS.368....2D, 2009Natur.457..451D}, or simply the continued expansion of the virial radius to a point where cooling from the halo to the disk becomes inefficient (i.e., too slow). 
While other environmental processes may also play a role --- for example, ram pressure, stripping, or harassment (see \citealt{2011nha..confE...8T} and \citealt{2013pss6.book..207V} for recent reviews) --- these local considerations cannot explain the global behavior of late-type galaxies as a class. 

The rate at  which star formation exhausts gas can be written in a dynamical form as \citep[][]{1998MNRAS.295..877G, 2003MNRAS.343...75H, 2011MNRAS.415.3798K}:

\begin{equation}
\psi = \frac{\epsilon M_{\rm gas}}{\tau_{\rm dyn}} , \label{eq:schmidt}
\end{equation}

\noindent where 
$\psi$ is the SFR, $M_{\rm gas}$ is the gas reservoir available for star formation, $\tau_{\rm dyn}$ is the dynamical time scale, and $\epsilon$ is the (fixed) efficiency of star formation \citep[we assume the universal value $\epsilon=2$\%; e.g.,][]{1998ApJ...498..541K}. The dynamical time scale for the disk of a massive galaxy is on the order of $\tau_{\rm dyn} \sim 50-300$ Myr. The mass of the gas reservoir is a free parameter set by the initial conditions. Once the initial gas reservoir is specified and a reasonable dynamical time scale is chosen, Eqn.~\ref{eq:schmidt} can be solved to describe the evolution of the system. 
The e-folding time for the declining star formation rate is $\tau_{\rm dyn}/ \epsilon$, which is several Gyr for the adopted parameters\footnote{This time scale for gas depletion corresponds to what happens in the ``bathtub'' model of \cite{2013arXiv1303.5059L} when the gas supply to the reservoir is shut off.}. That is, galaxies with a gas reservoir that does not get replenished from outside will continue to form stars for a very long time, peeling off from the main sequence of star formation very slowly, as previously argued by \cite{2011MNRAS.415.3798K}. In Figure \ref{fig:tracks}, we show how the specific star formation rates, $u-r$ colours, and gas fractions evolve according to Eqn.~3 for a range of quenching time scales. 

The evolution of late-type galaxies corresponds to the longer time scale tracks in Figure \ref{fig:tracks}. 
This means there is enough time to develop a significant population of late types in the green valley and, eventually, the red sequence. 
Late-type galaxies in the green valley should still have substantial gas reservoirs that fuel ongoing star formation. Initial data from HI data support this (Section \ref{sec:HI_gas}).

The start of quenching may occur at (very) high redshift, as it can take several Gyrs for the effects of a slowly declining star formation rate to become apparent. In this context, red spirals \citep[e.g.,][]{1976ApJ...206..883V, 2009MNRAS.393.1302W, 2010MNRAS.405..783M} are simply late-type galaxies whose cutoff happened relatively early. The observation of \cite{2010MNRAS.405..783M} and \cite{2012A&A...543A.132C} that red spirals still show signs of low-level star formation make sense --- they are simply very far along the exponential decline in SFR off the main sequence. 

The reddest colours in late-type galaxies preferentially occur in high mass haloes (Fig.~\ref{fig:gv_env}), supporting the suggestion that the gas supply is somehow regulated by environmental factors.  Of course, other factors besides halo mass can increase the pace of gas depletion following the original quenching event --- and this must happen: Figure \ref{fig:evol1} shows some objects with quenching time scales slightly shorter than that expected from canonical parameters. 

For example, bars naturally shorten gas consumption time scales, by driving gas into the central regions of galaxies where it can be consumed more quickly. \cite{2011MNRAS.411.2026M, 2012MNRAS.424.2180M} found that the incidence of bars in massive late-type galaxies increases as they become redder. This could be a side effect of quenching --- models show that bars form more quickly in gas-poor disks \citep[e.g.,][]{2013MNRAS.429.1949A} --- but if secular evolution along a bar is an efficient process, the presence of a bar could also speed up the quenching time such that the transition happens more rapidly than the several Gyrs expected from the simple Schmidt law. \cite{2012MNRAS.424.2180M} show that high bar fractions in spirals are associated with lower HI gas content, and there are hints that at a fixed HI content barred galaxies are optically redder than unbarred galaxies. Several works (\citealt{2008ApJ...675.1141S, 2010MNRAS.409..346C}; Melvin et al. in prep) show that the fraction of disk galaxies with bars also seems to have been increasing since $z\sim1$ \citet{2008ApJ...675.1141S, 2010MNRAS.409..346C}. Thus, secular processes like bar formation may be intimately tied to quenching in late-type galaxies; even if they appear after the initial quenching event, they would help drive out the remaining gas reservoir, thus accelerating the rate at which late types become fully gas- and star-formation free.

Environment as a late-stage accelerator or modifier of quenching is naturally supported by the strong effect of halo mass on the presence of green and red late types (Section \ref{sec:env_gv}). Other recent observational studies have argued that environment or satellite quenching should act fast, at least once it begins \citep[e.g.,][]{2012ApJ...746..188M, 2013MNRAS.432..336W}. 

Other environmental effects can accelerate the quenching process by further removing gas but not too quickly, as we do not see late-type galaxies with short quenching timescales. We note that these modulating effects (secular processes and environment) mean that most likely we cannot accurately reconstruct the time of quenching for present-day green and red (i.e., off-main sequence) late types.

Now we consider the possible role of AGN feedback in the evolution of late-type galaxies.The majority of black holes growing locally and at high redshift are hosted in late-type galaxies, clearly driven by secular processes rather than major mergers \citep[e.g.,][]{2010ApJ...711..284S,2011ApJ...727L..31S, 2012MNRAS.425L..61S, 2011ApJ...726...57C, 2012ApJ...744..148K, 2012ApJ...761...75S, 2012ApJ...758L..39T}. 
The fraction of green valley galaxies that host AGN is higher than in blue star-forming galaxies (Fig.~11). This could be due in part to an observational bias (the difficulty of seeing AGN signatures in SDSS spectra against a more luminous stellar component), but sensitive multiwavelength searches for  ``missed'' AGN in nearby star-forming galaxies \citep[e.g.,][]{2010MNRAS.406..597G} do not find enough to balance the numbers. 
Moreover, hard X-ray selection should not be confused by star formation, yet this technique also detects few to no (high luminosity) AGN in the blue cloud \citep{2009ApJ...692L..19S}. We conclude the prevalence of AGN in green valley galaxies is likely real rather than a selection effect.  

The alternative is that the onset of quenching in late-type galaxies leads to increased black hole growth about 1~Gyr later.  The delay could be explained naturally by the time required for accreting matter to lose angular momentum. In this case, black hole feeding would be due to secular processes (since major mergers would disrupt the disk morphology) and could not significantly speed up the quenching (or the AGN hosts would transition much more rapidly across $NUVur$ colour space); it would be a consequence of the host galaxy quenching, not the cause. Perhaps a declining sSFR favors black hole growth: \cite{2007ApJ...671.1388D} suggested that the absence of strong stellar feedback (core collapse supernovae, O-star winds) makes it easier to transport gas to the center, and also that mass loss from Asymptotic Giant Branch stars could easily fuel black hole accretion because of both the larger amount of material and its lower velocity. 

\cite{2010ApJ...721..193P} identified two distinct quenching mechanisms at work in galaxies: environment and mass quenching (without identifying the physical mechanisms). From the data presented in Section \ref{sec:env_gv}, one might speculate that early-type galaxies are mass quenched (no strong dependence on environment) and late types are environment quenched (strong dependence on environment). A test of this hypothesis would be the mass functions of green valley early- and late-type galaxies: if the quenching pathway we identify in early types is mass quenching
(see below), then the green valley early types should have the same mass function as the passive galaxies (in terms of $M*$ and faint end slope $\alpha_{\rm s}$, the normalisation $\Phi*$ will depend on transition time scale); similarly, if the quenching pathway we identify in late types is environment quenching, then the green valley late types should have the same mass function as the star-forming galaxies.

\subsection{The evolution of early-type galaxies} 

The end of star formation in early-type galaxies in the local Universe proceeds in a fundamentally different fashion than in the late-type galaxies. In local early types, quenching occurs in low-mass galaxies and is marked by a very rapid shutdown of star formation on time scales consistent with instantaneous suppression, or at most $\tau_{\rm quench} \sim 250$ Myr \citep[c.f.][]{2007MNRAS.382.1415S}. This rapid suppression is inconsistent with the simple gas exhaustion scenario outlined for late-type galaxies; the Schmidt law does not allow star formation to deplete a substantial gas reservoir so rapidly \citep[a point made strongly by][]{2011MNRAS.415.3798K}. The quenching mechanism should also be linked to the destruction of the disk morphology of the likely progenitor(s).

A major merger \citep{2010ApJ...714L.108S} could simultaneously transform the galaxy morphology from disk to spheroid and cause rapid depletion of the cold gas reservoir. Deep imaging of blue early types does reveal tidal features indicative of a major merger \citep{2010ApJ...714L.108S}
but \cite{2012MNRAS.420.1684W} have shown that galaxies with post-starburst spectral features already have early-type morphologies, 
emphasizing that the morphological transformation must be rapid.
 
Studies of gas kinematics (molecular, ionised) show that most early-type galaxies, and specifically those in the green valley, have gas with an external origin, most likely due to merger activity \citep{2006MNRAS.366.1151S,2010MNRAS.402.2140S,2013MNRAS.429..534D,2003ApJ...597L.117K}.
These blue precursors must have similar masses but bluer (intrinsic) colours than the AGN early-type hosts (see Figure \ref{fig:colour_mass_gv_nodust} and \citealt{2007MNRAS.382.1415S,2009MNRAS.396..818S}).  Observations of a small sample of early types along this evolutionary path from the blue cloud to the red sequence shows that the large molecular gas reservoirs of blue early-type galaxies do disappear rapidly,  at a rate significantly faster than can be explained by star formation alone \citep{2009ApJ...690.1672S}. 

We note that, as with the green valley late types, the quenching event --- i.e., the point at which the gas fuelling star formation was destroyed --- had to occur \textit{before} the galaxy reaches the green valley. Given the short time between quenching and arrival in the green valley (only a few hundred Megayears), it may be easier to identify this progenitor population than in late types. So far, \cite{2009ApJ...690.1672S} found that green valley (Seyfert AGN) early-type galaxies were undetected in molecular gas and that the gas disappeared rapidly during the transition from purely star-forming to an AGN+star formation mixed phase. 

Naturally, today's galaxies host both stellar populations formed {\it in situ} and those brought in via progenitors in minor and major mergers. In this general picture, star formation in early-type galaxies ceases and then does not re-start except for a minor frosting of new stellar populations \citep[seen mostly in the ultraviolet; e.g.,][]{2005ApJ...619L.111Y, 2007ApJS..173..512S, 2007ApJS..173..619K}, except when a new disk forms through the acquisition of a sufficient supply of cold gas, at which point the galaxy would rejoin the main sequence.

What about the role of AGN feedback? Rapid depletion of a large fraction of the available gas could be caused by a vigorous starburst, as star formation uses up cold gas and stellar processes create strong winds. Simulations show that merger-induced starbursts can lead to enhanced star formation as disk destruction leads to the inflow of gas to the (new) galaxy centre, all on relatively short dynamical time scales \citep[e.g.,][]{1996ApJ...471..115B}.  However, in this starburst-induced scenario, even a short depletion time scale will still yield a remnant system that is not entirely quenched. The Schmidt Law forces the star formation back to an exponentially decaying state, albeit with a shorter time scale, which in turn means that the galaxy retains gas, and therefore continues forming stars.  Only by adding AGN feedback, which can destroy the gas reservoir, do simulations show a genuine total quenching of star formation \citep[e.g.,][]{2005ApJ...620L..79S}.

Is it possible that AGN feedback alone destroys the gas reservoir during merger? It is true that photoionization happens almost instantaneously, so a luminous AGN could destroy the cold gas reservoir almost instantaneously. However, we would then expect detectable AGN radiation before the host galaxies reach the green valley, while they are still in the blue, star-forming phase, contrary to what is observed \citep[][]{2009ApJ...692L..19S, 2010MNRAS.406..597G}. If AGN feedback were responsible for the rapid gas reservoir destruction, then during this phase it must be either very short-lived, heavily obscured or radiatively inefficient \citep{2009ApJ...690.1672S,2012arXiv1206.2661S}.

A radiatively inefficient accretion flow could drive a kinetic outflow (jet, disk wind, or other outflow), analogous to what X-ray binaries do \citep[e.g.,][]{2003MNRAS.345L..19M,2006MNRAS.372.1366K, 2006Natur.444..730M, 2010Natur.466..209P}. In this early phase, the AGN would be outshone by the declining starburst; the place to look for evidence of this kind of kinetic feedback would be at the transition from star forming to composite spectrum in the blue early-type galaxy population.

We note that dwarf ellipticals are not in our sample and their quenching pathways may be very different \citep[e.g.][]{2008ApJ...674..742B}

%-----------------------------------------------------------------------------------------------------------------------------------
\begin{figure*}
\begin{center}

\includegraphics[width=0.99\textwidth]{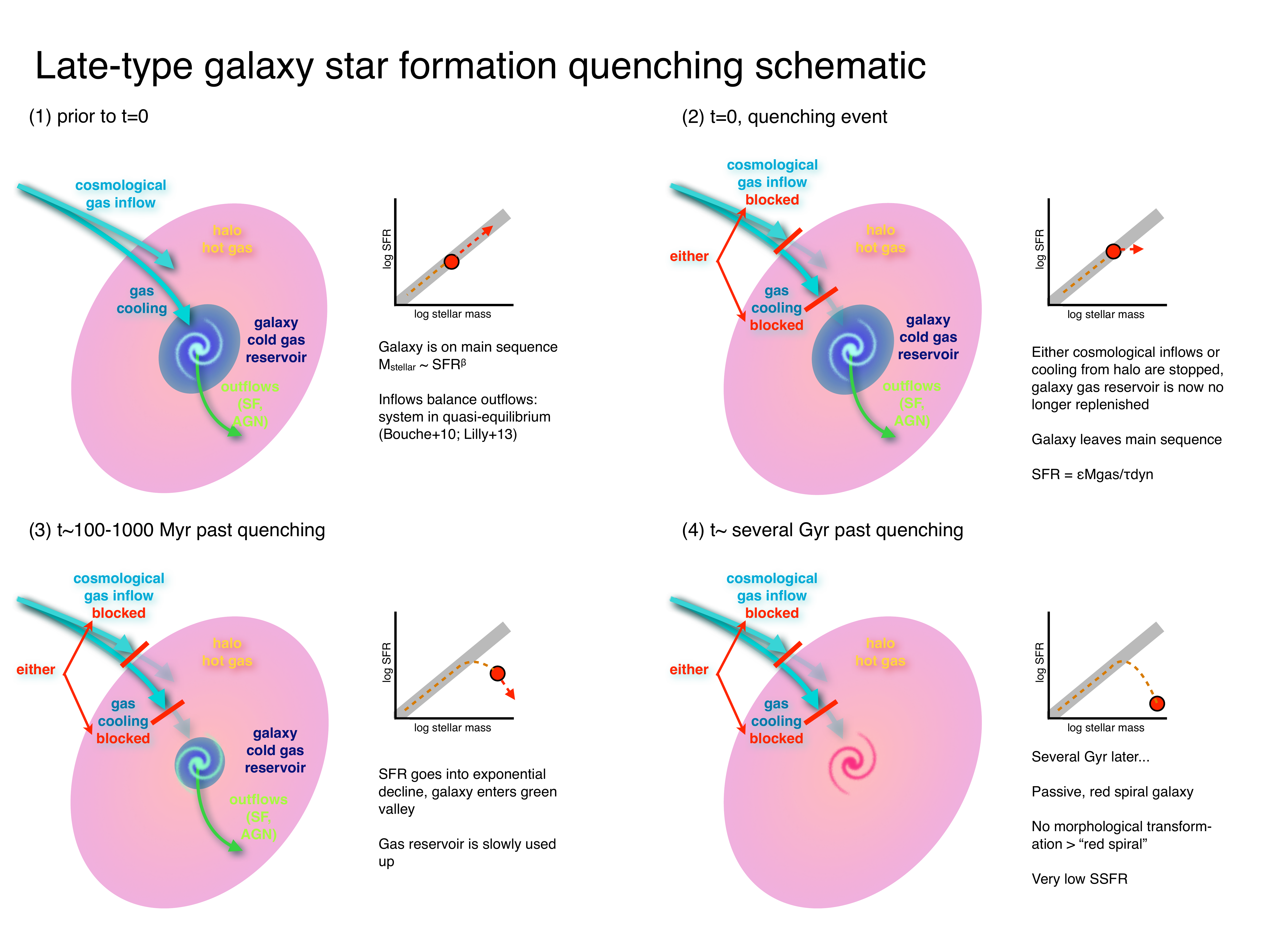}

\caption{Diagram outlining the scenario presented here for star-formation quenching in late-type galaxies. Once the external supply of gas to the galaxy's reservoir is cut off, the galaxy will leave the main sequence, and the SFR will decay exponentially, with a long quenching time scale $\tau_{\rm quench}$. While no longer on the main sequence, the galaxy nevertheless continues to convert gas to stars and thus to increase its stellar mass. Absent a major merger, this evolutionary pathway eventually  produces a passive, red, late-type galaxy. We see radiation from black hole growth during stage (3), long after the quenching event actually took place. }

\label{fig:lt_cartoon}

\end{center}
\end{figure*}
%-----------------------------------------------------------------------------------------------------------------------------------

%-----------------------------------------------------------------------------------------------------------------------------------
\begin{figure*}
\begin{center}

\includegraphics[width=0.99\textwidth]{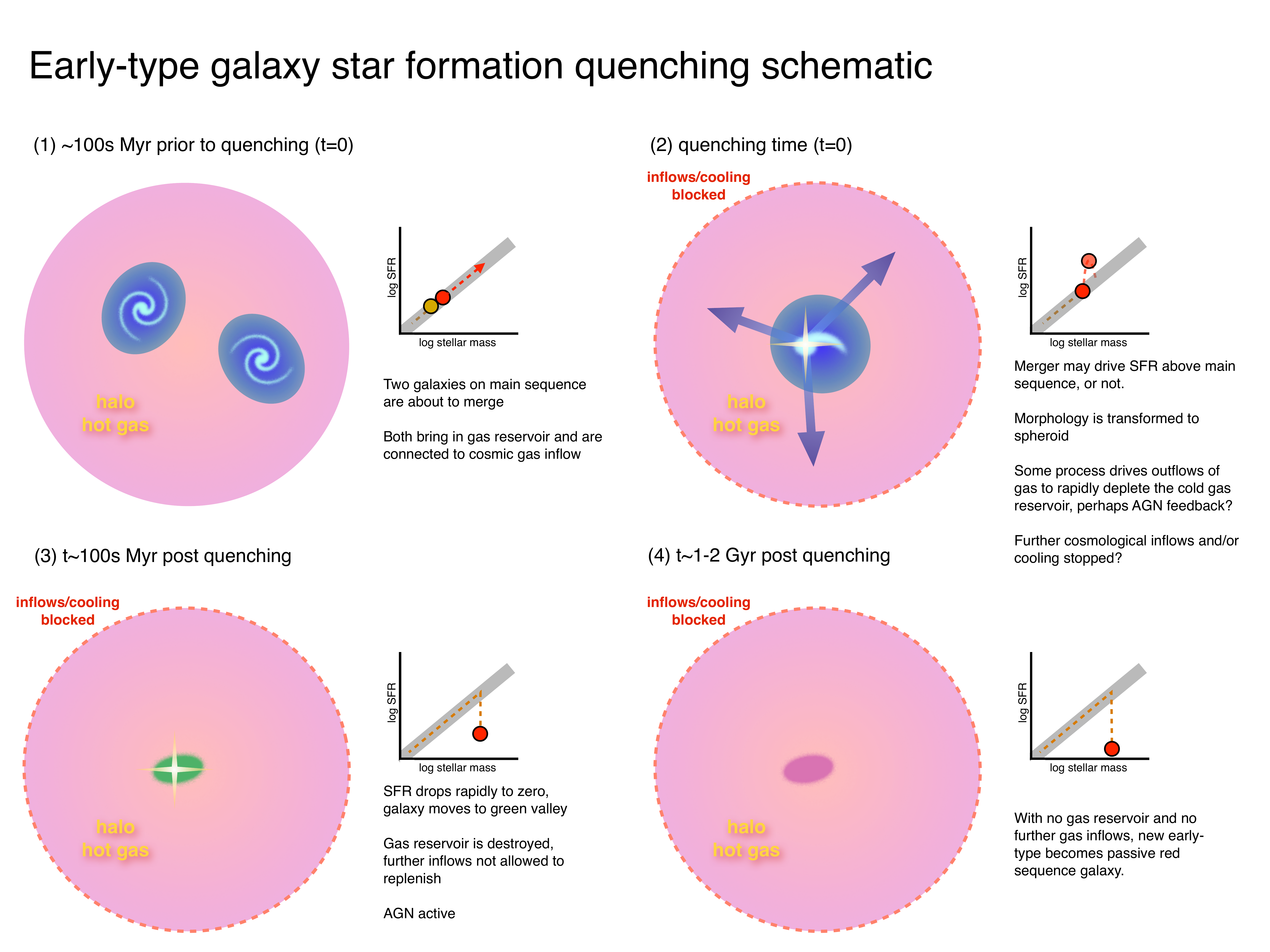}

\caption{Diagram outlining the scenario presented here for star-formation quenching in early-type galaxies. A dramatic event, likely a gas-rich merger of main sequence star-forming galaxies, causes the near-instantaneous destruction of the (newly formed) early-type galaxy gas reservoir, via stellar processes and/or AGN feedback. As long as the gas reservoir is not resupplied or the AGN keeps the gas hot, star formation ceases and no further stellar mass is formed. An AGN phase is visible for 100s of Myr as the stars in the early-type host galaxy age rapidly through the green valley. }

\label{fig:et_cartoon}

\end{center}
\end{figure*}
%-----------------------------------------------------------------------------------------------------------------------------------

%-----------------------------------------------------------------------------------------------------------------------------------
\section{Summary}
We have used new morphological analyses from Galaxy Zoo to map out the evolutionary pathways of local galaxies. We showed that the paths taken by late-type and early-type galaxies through the (optical) green valley are quite different, and that their quenching mechanisms have very different time scales. This means that thinking of the green valley as a transition phase for all (or even most) galaxies is overly simple. 
In particular, late-type galaxies do not exhibit the colour bimodality seen in the colour-mass or colour-magnitude diagrams of the total galaxy population.

From a detailed analysis of specific star formation rates, dust-corrected UV-optical colours and other properties, we traced the evolution of early- and late-type galaxies through the green valley. Both leave the main sequence and enter the green valley as their sSFR drops, but they do so with very different rates of change. UV-optical colours show that the rate of change in sSFR (d/dt sSFR) --- i.e., the quenching time scale --- is rapid in early-type galaxies ($\tau < 250$~Myr), while late-type galaxies undergo a much more gradual decline in star formation ($\tau > 1$~Gyr). We illustrate these morphology-related star formation quenching pathways with cartoons in Figures \ref{fig:lt_cartoon} and \ref{fig:et_cartoon}, as described here:\\

\noindent \textbf{Late-type galaxies:}

\begin{enumerate}

\item The quenching of star formation is initiated by a cutoff of the galaxy gas reservoir from the cosmic supply of fresh gas.  This cutoff could be due, for example, to the halo mass reaching $10^{12}$~$M_{\sun}$, preventing further accretion of gas onto the galaxy, or to cooling of the hot halo gas becoming inefficient.

\item This disturbance of the balance of inflows and outflows moves the galaxy off the main sequence, as star formation uses up the remaining gas and the gas reservoir is not replenished. Where initially the galaxy SFR scales nearly linearly with the stellar mass (the main sequence), it then declines exponentially after quenching commences, with a long characteristic time scale that to first order is set by the gas reservoir at the time quenching begins and the dynamical time scale of the galaxy disk.

\item 
Since the SFR is declining, but not zero, the stellar mass may continue to increase as star formation converts the remaining gas reservoir to stars.

\item
The galaxy moves slowly out of the blue cloud and into the green valley. Objects that were quenched this way at high redshift may by now have reached the red sequence, accounting for the red spiral population. 

\item
The gas-depletion process can be accelerated by other physical processes, in particular secular processes and environmental processes.

\item Black hole accretion appears to be favoured in late types that have been quenched and are in the exponential decline phase.

\item Observations of still-significant gas reservoirs and high dark matter halo masses support this evolutionary scenario.

\item The time delay between the quenching event (i.e., the point at which the external gas supply to the galaxy reservoir is cut off) and the time that the quenching becomes apparent (by movement out of the blue cloud and into the green valley) is long, on the order of several gigayears. This means that studying the local green valley galaxies will not allow us to understand this quenching mode or directly observe it in action. It also means that the green and red late types we see today may be the amongst the first to have quenched. The Milky Way may be on a similar trajectory to quiescence.
\end{enumerate}

\noindent \textbf{Early-type galaxies:}

\begin{enumerate}

\item The quenching of star formation is triggered by the rapid destruction of the galaxy gas reservoir. This must occur rapidly and can not be due to gas exhaustion by star formation alone. 

\item The destruction of the gas reservoir triggers the immediate departure from the main sequence of the galaxy. The SFR rapidly approaches zero, which means the galaxy no longer increases its stellar mass.

\item The drop in SFR corresponds to the galaxy moving out of the blue cloud, into the green valley and to the red sequence as fast as stellar evolution allows. The transition process in terms of galaxy colour takes about $1$~Gyr.

\item The rapid quenching event is effectively simultaneous with the morphological transformation, since there are very few blue early-type galaxies. This suggests a common origin in a major merger.

\item Visible radiation from black hole accretion is associated with the green valley, i.e., only \textit{after} the quenching event.

\item The rapidity of the gas reservoir destruction suggests that unusually strong stellar process and/or AGN feedback (winds, ionization) are involved, perhaps in a kinetic or highly obscured phase.

\item To understand the physics of quenching in early types more fully requires observations of the progenitors of the blue early types --- most likely major, gas-rich mergers --- where we can see which processes (AGN or not) destroy the gas reservoir.
\end{enumerate}

%-----------------------------------------------------------------------------------------------------------------------------------

%-----------------------------------------------------------------------------------------------------------------------------------

\section*{Acknowledgements}
We thank C. M. Carollo, S. Lilly, E. Bell and Y. Peng for useful discussions and A. Muench for help with the VO. We also thank the anonymous referee for helpful comments.

KS gratefully acknowledges support from Swiss National Science Foundation Grant PP00P2\_138979/1. BDS acknowledges support from the Oxford Martin School and Worcester College, Oxford. The development of Galaxy Zoo was supported by a Jim Gray grant from Microsoft and The Leverhulme Trust. RCN acknowledges STFC Rolling Grant ST/I001204/1 ``Survey Cosmology and Astrophysics''. SKY acknowledges the support from the National Research Foundation of Korea to the Center for Galaxy Evolution Research (No. 2010-0027910) and Korea Research Council of Fundamental Science and Technology (DRC Program, FY 2012). Support for the work of ET was provided by the Center of Excellence in Astrophysics and Associated Technologies (PFB 06), by the FONDECYT grant 1120061 and the Anillo project ACT1101. LFF and KWW acknowledge support the US National Science Foundation under grant DRL-0941610. RAS is supported by the NSF grant AST-1055081. SK acknowledges a Senior Research Fellowship from Worcester College, Oxford. All hail the Glow Cloud.

GALEX (Galaxy Evolution Explorer) is a NASA Small Explorer, launched in 2003 April. We gratefully acknowledge NASA's support for construction, operation, and science analysis for the GALEX mission, developed in cooperation with the Centre National d'Etudes Spatiales of France and the Korean Ministry of Science and Technology.

This publication makes use of data products from the Two Micron All Sky Survey, which is a joint project of the University of Massachusetts and the Infrared Processing and Analysis Center/California Institute of Technology, funded by the National Aeronautics and Space Administration and the National Science Foundation.

Funding for the SDSS and SDSS-II has been provided by the Alfred P. Sloan Foundation, the Participating Institutions, the National Science Foundation, the U.S. Department of Energy, the National Aeronautics and Space Administration, the Japanese Monbukagakusho, the Max Planck Society, and the Higher Education Funding Council for England. The SDSS Web Site is http://www.sdss.org/.

The SDSS is managed by the Astrophysical Research Consortium for the Participating Institutions. The Participating Institutions are the American Museum of Natural History, Astrophysical Institute Potsdam, University of Basel, University of Cambridge, Case Western Reserve University, University of Chicago, Drexel University, Fermilab, the Institute for Advanced Study, the Japan Participation Group, Johns Hopkins University, the Joint Institute for Nuclear Astrophysics, the Kavli Institute for Particle Astrophysics and Cosmology, the Korean Scientist Group, the Chinese Academy of Sciences (LAMOST), Los Alamos National Laboratory, the Max-Planck-Institute for Astronomy (MPIA), the Max-Planck-Institute for Astrophysics (MPA), New Mexico State University, Ohio State University, University of Pittsburgh, University of Portsmouth, Princeton University, the United States Naval Observatory, and the University of Washington.

This publication made extensive use of the Tool for OPerations on Catalogues And Tables (TOPCAT), which can be found at \texttt{http://www.starlink.ac.uk/topcat/}. This research has made use of NASA's ADS Service.
% Hello there, @OverheardOnAph

%-----------------------------------------------------------------------------------------------------------------------------------

\bibliographystyle{mn2e}
%\bibliography{bibliography2}

\begin{thebibliography}{122}
\expandafter\ifx\csname natexlab\endcsname\relax\def\natexlab#1{#1}\fi

\bibitem[{{Abazajian} {et~al}\mbox{.}(2009){Abazajian}, {Adelman-McCarthy},
  {Ag{\"u}eros}, {Allam}, {Allende Prieto}, {An}, {Anderson}, {Anderson},
  {Annis}, {Bahcall}, {Bailer-Jones}, {Barentine}, {Bassett}, {Becker},
  {Beers}, {Bell}, {Belokurov}, {Berlind}, {Berman}, {Bernardi}, {Bickerton},
  {Bizyaev}, {Blakeslee}, {Blanton}, {Bochanski}, {Boroski}, {Brewington},
  {Brinchmann}, {Brinkmann}, {Brunner}, {Budav{\'a}ri}, {Carey}, {Carliles},
  {Carr}, {Castander}, {Cinabro}, {Connolly}, {Csabai}, {Cunha}, {Czarapata},
  {Davenport}, {de Haas}, {Dilday}, {Doi}, {Eisenstein}, {Evans}, {Evans},
  {Fan}, {Friedman}, {Frieman}, {Fukugita}, {G{\"a}nsicke}, {Gates},
  {Gillespie}, {Gilmore}, {Gonzalez}, {Gonzalez}, {Grebel}, {Gunn},
  {Gy{\"o}ry}, {Hall}, {Harding}, {Harris}, {Harvanek}, {Hawley}, {Hayes},
  {Heckman}, {Hendry}, {Hennessy}, {Hindsley}, {Hoblitt}, {Hogan}, {Hogg},
  {Holtzman}, {Hyde}, {Ichikawa}, {Ichikawa}, {Im}, {Ivezi{\'c}}, {Jester},
  {Jiang}, {Johnson}, {Jorgensen}, {Juri{\'c}}, {Kent}, {Kessler}, {Kleinman},
  {Knapp}, {Konishi}, {Kron}, {Krzesinski}, {Kuropatkin}, {Lampeitl},
  {Lebedeva}, {Lee}, {Lee}, {Leger}, {L{\'e}pine}, {Li}, {Lima}, {Lin}, {Long},
  {Loomis}, {Loveday}, {Lupton}, {Magnier}, {Malanushenko}, {Malanushenko},
  {Mandelbaum}, {Margon}, {Marriner}, {Mart{\'{\i}}nez-Delgado}, {Matsubara},
  {McGehee}, {McKay}, {Meiksin}, {Morrison}, {Mullally}, {Munn}, {Murphy},
  {Nash}, {Nebot}, {Neilsen}, {Newberg}, {Newman}, {Nichol}, {Nicinski},
  {Nieto-Santisteban}, {Nitta}, {Okamura}, {Oravetz}, {Ostriker}, {Owen},
  {Padmanabhan}, {Pan}, {Park}, {Pauls}, {Peoples}, {Percival}, {Pier}, {Pope},
  {Pourbaix}, {Price}, {Purger}, {Quinn}, {Raddick}, {Fiorentin}, {Richards},
  {Richmond}, {Riess}, {Rix}, {Rockosi}, {Sako}, {Schlegel}, {Schneider},
  {Scholz}, {Schreiber}, {Schwope}, {Seljak}, {Sesar}, {Sheldon}, {Shimasaku},
  {Sibley}, {Simmons}, {Sivarani}, {Smith}, {Smith}, {Smol{\v c}i{\'c}},
  {Snedden}, {Stebbins}, {Steinmetz}, {Stoughton}, {Strauss}, {Subba Rao},
  {Suto}, {Szalay}, {Szapudi}, {Szkody}, {Tanaka}, {Tegmark}, {Teodoro},
  {Thakar}, {Tremonti}, {Tucker}, {Uomoto}, {Vanden Berk}, {Vandenberg},
  {Vidrih}, {Vogeley}, {Voges}, {Vogt}, {Wadadekar}, {Watters}, {Weinberg},
  {West}, {White}, {Wilhite}, {Wonders}, {Yanny}, {Yocum}, {York}, {Zehavi},
  {Zibetti}, \& {Zucker}}]{2009ApJS..182..543A}
{Abazajian} K.~N. {et~al.}, 2009, \apjs, 182, 543

\bibitem[{{Athanassoula} {et~al}\mbox{.}(2013){Athanassoula}, {Machado}, \&
  {Rodionov}}]{2013MNRAS.429.1949A}
{Athanassoula} E., {Machado} R.~E.~G., {Rodionov} S.~A., 2013, \mnras, 429,
  1949

\bibitem[{{Baldry} {et~al}\mbox{.}(2006){Baldry}, {Balogh}, {Bower},
  {Glazebrook}, {Nichol}, {Bamford}, \& {Budavari}}]{2006MNRAS.373..469B}
{Baldry} I.~K., {Balogh} M.~L., {Bower} R.~G., {Glazebrook} K., {Nichol} R.~C.,
  {Bamford} S.~P., {Budavari} T., 2006, \mnras, 373, 469

\bibitem[{{Baldry} {et~al}\mbox{.}(2004){Baldry}, {Glazebrook}, {Brinkmann},
  {Ivezi{\'c}}, {Lupton}, {Nichol}, \& {Szalay}}]{2004ApJ...600..681B}
{Baldry} I.~K., {Glazebrook} K., {Brinkmann} J., {Ivezi{\'c}} {\v Z}., {Lupton}
  R.~H., {Nichol} R.~C., {Szalay} A.~S., 2004, \apj, 600, 681

\bibitem[{{Bamford} {et~al}\mbox{.}(2009){Bamford}, {Nichol}, {Baldry}, {Land},
  {Lintott}, {Schawinski}, {Slosar}, {Szalay}, {Thomas}, {Torki}, {Andreescu},
  {Edmondson}, {Miller}, {Murray}, {Raddick}, \&
  {Vandenberg}}]{2009MNRAS.393.1324B}
{Bamford} S.~P. {et~al.}, 2009, \mnras, 393, 1324

\bibitem[{{Barnes} \& {Hernquist}(1996)}]{1996ApJ...471..115B}
{Barnes} J.~E., {Hernquist} L., 1996, \apj, 471, 115

\bibitem[{{Behroozi} {et~al}\mbox{.}(2010){Behroozi}, {Conroy}, \&
  {Wechsler}}]{2010ApJ...717..379B}
{Behroozi} P.~S., {Conroy} C., {Wechsler} R.~H., 2010, \apj, 717, 379

\bibitem[{{Bell} {et~al}\mbox{.}(2004){Bell}, {Wolf}, {Meisenheimer}, {Rix},
  {Borch}, {Dye}, {Kleinheinrich}, {Wisotzki}, \&
  {McIntosh}}]{2004ApJ...608..752B}
{Bell} E.~F. {et~al.}, 2004, \apj, 608, 752

\bibitem[{{Blanton} \& {Roweis}(2007)}]{2007AJ....133..734B}
{Blanton} M.~R., {Roweis} S., 2007, \aj, 133, 734

\bibitem[{{Blanton} {et~al}\mbox{.}(2005){Blanton}, {Schlegel}, {Strauss},
  {Brinkmann}, {Finkbeiner}, {Fukugita}, {Gunn}, {Hogg}, {Ivezi{\'c}}, {Knapp},
  {Lupton}, {Munn}, {Schneider}, {Tegmark}, \& {Zehavi}}]{2005AJ....129.2562B}
{Blanton} M.~R. {et~al.}, 2005, \aj, 129, 2562

\bibitem[{{Boselli} {et~al}\mbox{.}(2008){Boselli}, {Boissier}, {Cortese}, \&
  {Gavazzi}}]{2008ApJ...674..742B}
{Boselli} A., {Boissier} S., {Cortese} L., {Gavazzi} G., 2008, \apj, 674, 742

\bibitem[{{Bouch{\'e}} {et~al}\mbox{.}(2010){Bouch{\'e}}, {Dekel}, {Genzel},
  {Genel}, {Cresci}, {F{\"o}rster Schreiber}, {Shapiro}, {Davies}, \&
  {Tacconi}}]{2010ApJ...718.1001B}
{Bouch{\'e}} N. {et~al.}, 2010, \apj, 718, 1001

\bibitem[{{Brammer} {et~al}\mbox{.}(2009){Brammer}, {Whitaker}, {van Dokkum},
  {Marchesini}, {Labb{\'e}}, {Franx}, {Kriek}, {Quadri}, {Illingworth}, {Lee},
  {Muzzin}, \& {Rudnick}}]{2009ApJ...706L.173B}
{Brammer} G.~B. {et~al.}, 2009, \apjl, 706, L173

\bibitem[{{Brinchmann} {et~al}\mbox{.}(2004){Brinchmann}, {Charlot}, {White},
  {Tremonti}, {Kauffmann}, {Heckman}, \& {Brinkmann}}]{2004MNRAS.351.1151B}
{Brinchmann} J., {Charlot} S., {White} S.~D.~M., {Tremonti} C., {Kauffmann} G.,
  {Heckman} T., {Brinkmann} J., 2004, \mnras, 351, 1151

\bibitem[{{Bruzual} \& {Charlot}(2003)}]{2003MNRAS.344.1000B}
{Bruzual} G., {Charlot} S., 2003, \mnras, 344, 1000

\bibitem[{{Calzetti} {et~al}\mbox{.}(2000){Calzetti}, {Armus}, {Bohlin},
  {Kinney}, {Koornneef}, \& {Storchi-Bergmann}}]{2000ApJ...533..682C}
{Calzetti} D., {Armus} L., {Bohlin} R.~C., {Kinney} A.~L., {Koornneef} J.,
  {Storchi-Bergmann} T., 2000, \apj, 533, 682

\bibitem[{{Cameron} {et~al}\mbox{.}(2010){Cameron}, {Carollo}, {Oesch},
  {Aller}, {Bschorr}, {Cerulo}, {Aussel}, {Capak}, {Le Floc'h}, {Ilbert},
  {Kneib}, {Koekemoer}, {Leauthaud}, {Lilly}, {Massey}, {McCracken}, {Rhodes},
  {Salvato}, {Sanders}, {Scoville}, {Sheth}, {Taniguchi}, \&
  {Thompson}}]{2010MNRAS.409..346C}
{Cameron} E. {et~al.}, 2010, \mnras, 409, 346

\bibitem[{{Cappellari} \& {Emsellem}(2004)}]{2004PASP..116..138C}
{Cappellari} M., {Emsellem} E., 2004, \pasp, 116, 138

\bibitem[{{Cardamone} {et~al}\mbox{.}(2009){Cardamone}, {Schawinski}, {Sarzi},
  {Bamford}, {Bennert}, {Urry}, {Lintott}, {Keel}, {Parejko}, {Nichol},
  {Thomas}, {Andreescu}, {Murray}, {Raddick}, {Slosar}, {Szalay}, \&
  {Vandenberg}}]{2009MNRAS.399.1191C}
{Cardamone} C. {et~al.}, 2009, \mnras, 399, 1191

\bibitem[{{Cardamone} {et~al}\mbox{.}(2010){Cardamone}, {Urry}, {Schawinski},
  {Treister}, {Brammer}, \& {Gawiser}}]{2010ApJ...721L..38C}
{Cardamone} C.~N., {Urry} C.~M., {Schawinski} K., {Treister} E., {Brammer} G.,
  {Gawiser} E., 2010, \apjl, 721, L38

\bibitem[{{Cardelli} {et~al}\mbox{.}(1989){Cardelli}, {Clayton}, \&
  {Mathis}}]{1989ApJ...345..245C}
{Cardelli} J.~A., {Clayton} G.~C., {Mathis} J.~S., 1989, \apj, 345, 245

\bibitem[{{Carollo} {et~al}\mbox{.}(2012){Carollo}, {Cibinel}, {Lilly},
  {Miniati}, {Norberg}, {Silverman}, {van Gorkom}, {Cameron}, {Finoguenov},
  {Pipino}, {Rudick}, {Lu}, \& {Peng}}]{2012arXiv1206.5807C}
{Carollo} C.~M. {et~al.}, 2012, ArXiv e-prints, 1206.5807

\bibitem[{{Cattaneo} {et~al}\mbox{.}(2006){Cattaneo}, {Dekel}, {Devriendt},
  {Guiderdoni}, \& {Blaizot}}]{2006MNRAS.370.1651C}
{Cattaneo} A., {Dekel} A., {Devriendt} J., {Guiderdoni} B., {Blaizot} J., 2006,
  \mnras, 370, 1651

\bibitem[{{Cheung} {et~al}\mbox{.}(2013){Cheung}, {Athanassoula}, {Masters},
  {Nichol}, {Bosma}, {Bell}, {Faber}, {Koo}, {Lintott}, {Melvin}, {Schawinski},
  {Skibba}, \& {Willett}}]{2013ApJ...779..162C}
{Cheung} E. {et~al.}, 2013, \apj, 779, 162

\bibitem[{{Cibinel} {et~al}\mbox{.}(2012){Cibinel}, {Carollo}, {Lilly},
  {Bonoli}, {Miniati}, {Pipino}, {Silverman}, {van Gorkom}, {Cameron},
  {Finoguenov}, {Norberg}, {Rudick}, {Lu}, \& {Peng}}]{2012arXiv1206.6496C}
{Cibinel} A. {et~al.}, 2012, ArXiv e-prints, 1206.6496

\bibitem[{{Cisternas} {et~al}\mbox{.}(2011){Cisternas}, {Jahnke}, {Inskip},
  {Kartaltepe}, {Koekemoer}, {Lisker}, {Robaina}, {Scodeggio}, {Sheth},
  {Trump}, {Andrae}, {Miyaji}, {Lusso}, {Brusa}, {Capak}, {Cappelluti},
  {Civano}, {Ilbert}, {Impey}, {Leauthaud}, {Lilly}, {Salvato}, {Scoville}, \&
  {Taniguchi}}]{2011ApJ...726...57C}
{Cisternas} M. {et~al.}, 2011, \apj, 726, 57

\bibitem[{{Cortese}(2012)}]{2012A&A...543A.132C}
{Cortese} L., 2012, \aap, 543, A132

\bibitem[{{Darg} {et~al}\mbox{.}(2010{\natexlab{a}}){Darg}, {Kaviraj},
  {Lintott}, {Schawinski}, {Sarzi}, {Bamford}, {Silk}, {Andreescu}, {Murray},
  {Nichol}, {Raddick}, {Slosar}, {Szalay}, {Thomas}, \&
  {Vandenberg}}]{2010MNRAS.401.1552D}
{Darg} D.~W. {et~al.}, 2010{\natexlab{a}}, \mnras, 401, 1552

\bibitem[{{Darg} {et~al}\mbox{.}(2010{\natexlab{b}}){Darg}, {Kaviraj},
  {Lintott}, {Schawinski}, {Sarzi}, {Bamford}, {Silk}, {Proctor}, {Andreescu},
  {Murray}, {Nichol}, {Raddick}, {Slosar}, {Szalay}, {Thomas}, \&
  {Vandenberg}}]{2010MNRAS.401.1043D}
{Darg} D.~W. {et~al.}, 2010{\natexlab{b}}, \mnras, 401, 1043

\bibitem[{{Davies} {et~al}\mbox{.}(2007){Davies}, {M{\"u}ller S{\'a}nchez},
  {Genzel}, {Tacconi}, {Hicks}, {Friedrich}, \&
  {Sternberg}}]{2007ApJ...671.1388D}
{Davies} R.~I., {M{\"u}ller S{\'a}nchez} F., {Genzel} R., {Tacconi} L.~J.,
  {Hicks} E.~K.~S., {Friedrich} S., {Sternberg} A., 2007, \apj, 671, 1388

\bibitem[{{Davis} {et~al}\mbox{.}(2013){Davis}, {Alatalo}, {Bureau},
  {Cappellari}, {Scott}, {Young}, {Blitz}, {Crocker}, {Bayet}, {Bois},
  {Bournaud}, {Davies}, {de Zeeuw}, {Duc}, {Emsellem}, {Khochfar},
  {Krajnovi{\'c}}, {Kuntschner}, {Lablanche}, {McDermid}, {Morganti}, {Naab},
  {Oosterloo}, {Sarzi}, {Serra}, \& {Weijmans}}]{2013MNRAS.429..534D}
{Davis} T.~A. {et~al.}, 2013, \mnras, 429, 534

\bibitem[{{Dekel} \& {Birnboim}(2006)}]{2006MNRAS.368....2D}
{Dekel} A., {Birnboim} Y., 2006, \mnras, 368, 2

\bibitem[{{Dekel} {et~al}\mbox{.}(2009){Dekel}, {Birnboim}, {Engel},
  {Freundlich}, {Goerdt}, {Mumcuoglu}, {Neistein}, {Pichon}, {Teyssier}, \&
  {Zinger}}]{2009Natur.457..451D}
{Dekel} A. {et~al.}, 2009, \nat, 457, 451

\bibitem[{{Elbaz} {et~al}\mbox{.}(2007){Elbaz}, {Daddi}, {Le Borgne},
  {Dickinson}, {Alexander}, {Chary}, {Starck}, {Brandt}, {Kitzbichler},
  {MacDonald}, {Nonino}, {Popesso}, {Stern}, \&
  {Vanzella}}]{2007A&A...468...33E}
{Elbaz} D. {et~al.}, 2007, \aap, 468, 33

\bibitem[{{Elbaz} {et~al}\mbox{.}(2011){Elbaz}, {Dickinson}, {Hwang},
  {D{\'{\i}}az-Santos}, {Magdis}, {Magnelli}, {Le Borgne}, {Galliano},
  {Pannella}, {Chanial}, {Armus}, {Charmandaris}, {Daddi}, {Aussel}, {Popesso},
  {Kartaltepe}, {Altieri}, {Valtchanov}, {Coia}, {Dannerbauer}, {Dasyra},
  {Leiton}, {Mazzarella}, {Alexander}, {Buat}, {Burgarella}, {Chary}, {Gilli},
  {Ivison}, {Juneau}, {Le Floc'h}, {Lutz}, {Morrison}, {Mullaney}, {Murphy},
  {Pope}, {Scott}, {Brodwin}, {Calzetti}, {Cesarsky}, {Charlot}, {Dole},
  {Eisenhardt}, {Ferguson}, {F{\"o}rster Schreiber}, {Frayer}, {Giavalisco},
  {Huynh}, {Koekemoer}, {Papovich}, {Reddy}, {Surace}, {Teplitz}, {Yun}, \&
  {Wilson}}]{2011A&A...533A.119E}
{Elbaz} D. {et~al.}, 2011, \aap, 533, A119

\bibitem[{{Faber} {et~al}\mbox{.}(2007){Faber}, {Willmer}, {Wolf}, {Koo},
  {Weiner}, {Newman}, {Im}, {Coil}, {Conroy}, {Cooper}, {Davis}, {Finkbeiner},
  {Gerke}, {Gebhardt}, {Groth}, {Guhathakurta}, {Harker}, {Kaiser}, {Kassin},
  {Kleinheinrich}, {Konidaris}, {Kron}, {Lin}, {Luppino}, {Madgwick},
  {Meisenheimer}, {Noeske}, {Phillips}, {Sarajedini}, {Schiavon}, {Simard},
  {Szalay}, {Vogt}, \& {Yan}}]{2007ApJ...665..265F}
{Faber} S.~M. {et~al.}, 2007, \apj, 665, 265

\bibitem[{{Gon{\c c}alves} {et~al}\mbox{.}(2012){Gon{\c c}alves}, {Martin},
  {Men{\'e}ndez-Delmestre}, {Wyder}, \& {Koekemoer}}]{2012ApJ...759...67G}
{Gon{\c c}alves} T.~S., {Martin} D.~C., {Men{\'e}ndez-Delmestre} K., {Wyder}
  T.~K., {Koekemoer} A., 2012, \apj, 759, 67

\bibitem[{{Goulding} {et~al}\mbox{.}(2010){Goulding}, {Alexander}, {Lehmer}, \&
  {Mullaney}}]{2010MNRAS.406..597G}
{Goulding} A.~D., {Alexander} D.~M., {Lehmer} B.~D., {Mullaney} J.~R., 2010,
  \mnras, 406, 597

\bibitem[{{Guiderdoni} {et~al}\mbox{.}(1998){Guiderdoni}, {Hivon}, {Bouchet},
  \& {Maffei}}]{1998MNRAS.295..877G}
{Guiderdoni} B., {Hivon} E., {Bouchet} F.~R., {Maffei} B., 1998, \mnras, 295,
  877

\bibitem[{{Hasinger}(2008)}]{2008A&A...490..905H}
{Hasinger} G., 2008, \aap, 490, 905

\bibitem[{{Hatton} {et~al}\mbox{.}(2003){Hatton}, {Devriendt}, {Ninin},
  {Bouchet}, {Guiderdoni}, \& {Vibert}}]{2003MNRAS.343...75H}
{Hatton} S., {Devriendt} J.~E.~G., {Ninin} S., {Bouchet} F.~R., {Guiderdoni}
  B., {Vibert} D., 2003, \mnras, 343, 75

\bibitem[{{Haynes} {et~al}\mbox{.}(2011){Haynes}, {Giovanelli}, {Martin},
  {Hess}, {Saintonge}, {Adams}, {Hallenbeck}, {Hoffman}, {Huang}, {Kent},
  {Koopmann}, {Papastergis}, {Stierwalt}, {Balonek}, {Craig}, {Higdon},
  {Kornreich}, {Miller}, {O'Donoghue}, {Olowin}, {Rosenberg}, {Spekkens},
  {Troischt}, \& {Wilcots}}]{2011AJ....142..170H}
{Haynes} M.~P. {et~al.}, 2011, \aj, 142, 170

\bibitem[{{Hoyle} {et~al}\mbox{.}(2011){Hoyle}, {Masters}, {Nichol},
  {Edmondson}, {Smith}, {Lintott}, {Scranton}, {Bamford}, {Schawinski}, \&
  {Thomas}}]{2011MNRAS.415.3627H}
{Hoyle} B. {et~al.}, 2011, \mnras, 415, 3627

\bibitem[{{Kauffmann} {et~al}\mbox{.}(2003){Kauffmann}, {Heckman}, {White},
  {Charlot}, {Tremonti}, {Brinchmann}, {Bruzual}, {Peng}, {Seibert},
  {Bernardi}, {Blanton}, {Brinkmann}, {Castander}, {Cs{\'a}bai}, {Fukugita},
  {Ivezic}, {Munn}, {Nichol}, {Padmanabhan}, {Thakar}, {Weinberg}, \&
  {York}}]{2003MNRAS.341...33K}
{Kauffmann} G. {et~al.}, 2003, \mnras, 341, 33

\bibitem[{{Kaviraj} {et~al}\mbox{.}(2012){Kaviraj}, {Darg}, {Lintott},
  {Schawinski}, \& {Silk}}]{2012MNRAS.419...70K}
{Kaviraj} S., {Darg} D., {Lintott} C., {Schawinski} K., {Silk} J., 2012,
  \mnras, 419, 70

\bibitem[{{Kaviraj} {et~al}\mbox{.}(2007){Kaviraj}, {Schawinski}, {Devriendt},
  {Ferreras}, {Khochfar}, {Yoon}, {Yi}, {Deharveng}, {Boselli}, {Barlow},
  {Conrow}, {Forster}, {Friedman}, {Martin}, {Morrissey}, {Neff},
  {Schiminovich}, {Seibert}, {Small}, {Wyder}, {Bianchi}, {Donas}, {Heckman},
  {Lee}, {Madore}, {Milliard}, {Rich}, \& {Szalay}}]{2007ApJS..173..619K}
{Kaviraj} S. {et~al.}, 2007, \apjs, 173, 619

\bibitem[{{Kaviraj} {et~al}\mbox{.}(2011){Kaviraj}, {Schawinski}, {Silk}, \&
  {Shabala}}]{2011MNRAS.415.3798K}
{Kaviraj} S., {Schawinski} K., {Silk} J., {Shabala} S.~S., 2011, \mnras, 415,
  3798

\bibitem[{{Keel} {et~al}\mbox{.}(2012){Keel}, {Chojnowski}, {Bennert},
  {Schawinski}, {Lintott}, {Lynn}, {Pancoast}, {Harris}, {Nierenberg},
  {Sonnenfeld}, \& {Proctor}}]{2012MNRAS.420..878K}
{Keel} W.~C. {et~al.}, 2012, \mnras, 420, 878

\bibitem[{{Kennicutt}(1998)}]{1998ApJ...498..541K}
{Kennicutt}, Jr. R.~C., 1998, \apj, 498, 541

\bibitem[{{Khochfar} \& {Burkert}(2003)}]{2003ApJ...597L.117K}
{Khochfar} S., {Burkert} A., 2003, \apjl, 597, L117

\bibitem[{{Kocevski} {et~al}\mbox{.}(2012){Kocevski}, {Faber}, {Mozena},
  {Koekemoer}, {Nandra}, {Rangel}, {Laird}, {Brusa}, {Wuyts}, {Trump}, {Koo},
  {Somerville}, {Bell}, {Lotz}, {Alexander}, {Bournaud}, {Conselice}, {Dahlen},
  {Dekel}, {Donley}, {Dunlop}, {Finoguenov}, {Georgakakis}, {Giavalisco},
  {Guo}, {Grogin}, {Hathi}, {Juneau}, {Kartaltepe}, {Lucas}, {McGrath},
  {McIntosh}, {Mobasher}, {Robaina}, {Rosario}, {Straughn}, {van der Wel}, \&
  {Villforth}}]{2012ApJ...744..148K}
{Kocevski} D.~D. {et~al.}, 2012, \apj, 744, 148

\bibitem[{{Komatsu} {et~al}\mbox{.}(2011){Komatsu}, {Smith}, {Dunkley},
  {Bennett}, {Gold}, {Hinshaw}, {Jarosik}, {Larson}, {Nolta}, {Page},
  {Spergel}, {Halpern}, {Hill}, {Kogut}, {Limon}, {Meyer}, {Odegard}, {Tucker},
  {Weiland}, {Wollack}, \& {Wright}}]{2011ApJS..192...18K}
{Komatsu} E. {et~al.}, 2011, \apjs, 192, 18

\bibitem[{{K{\"o}rding} {et~al}\mbox{.}(2006){K{\"o}rding}, {Jester}, \&
  {Fender}}]{2006MNRAS.372.1366K}
{K{\"o}rding} E.~G., {Jester} S., {Fender} R., 2006, \mnras, 372, 1366

\bibitem[{{Land} {et~al}\mbox{.}(2008){Land}, {Slosar}, {Lintott}, {Andreescu},
  {Bamford}, {Murray}, {Nichol}, {Raddick}, {Schawinski}, {Szalay}, {Thomas},
  \& {Vandenberg}}]{2008MNRAS.388.1686L}
{Land} K. {et~al.}, 2008, \mnras, 388, 1686

\bibitem[{{Lee} {et~al}\mbox{.}(2012){Lee}, {Ferguson}, {Wiklind}, {Dahlen},
  {Dickinson}, {Giavalisco}, {Grogin}, {Papovich}, {Messias}, {Guo}, \&
  {Lin}}]{2012ApJ...752...66L}
{Lee} K.-S. {et~al.}, 2012, \apj, 752, 66

\bibitem[{{Leitner}(2012)}]{2012ApJ...745..149L}
{Leitner} S.~N., 2012, \apj, 745, 149

\bibitem[{{Lilly} {et~al}\mbox{.}(2013){Lilly}, {Carollo}, {Pipino}, {Renzini},
  \& {Peng}}]{2013arXiv1303.5059L}
{Lilly} S.~J., {Carollo} C.~M., {Pipino} A., {Renzini} A., {Peng} Y., 2013,
  ArXiv e-prints, 1303.5059

\bibitem[{{Lintott} {et~al}\mbox{.}(2011){Lintott}, {Schawinski}, {Bamford},
  {Slosar}, {Land}, {Thomas}, {Edmondson}, {Masters}, {Nichol}, {Raddick},
  {Szalay}, {Andreescu}, {Murray}, \& {Vandenberg}}]{2011MNRAS.410..166L}
{Lintott} C. {et~al.}, 2011, \mnras, 410, 166

\bibitem[{{Lintott} {et~al}\mbox{.}(2009){Lintott}, {Schawinski}, {Keel}, {van
  Arkel}, {Bennert}, {Edmondson}, {Thomas}, {Smith}, {Herbert}, {Jarvis},
  {Virani}, {Andreescu}, {Bamford}, {Land}, {Murray}, {Nichol}, {Raddick},
  {Slosar}, {Szalay}, \& {Vandenberg}}]{2009MNRAS.399..129L}
{Lintott} C.~J. {et~al.}, 2009, \mnras, 399, 129

\bibitem[{{Lintott} {et~al}\mbox{.}(2008){Lintott}, {Schawinski}, {Slosar},
  {Land}, {Bamford}, {Thomas}, {Raddick}, {Nichol}, {Szalay}, {Andreescu},
  {Murray}, \& {Vandenberg}}]{2008MNRAS.389.1179L}
{Lintott} C.~J. {et~al.}, 2008, \mnras, 389, 1179

\bibitem[{{Maccarone} {et~al}\mbox{.}(2003){Maccarone}, {Gallo}, \&
  {Fender}}]{2003MNRAS.345L..19M}
{Maccarone} T.~J., {Gallo} E., {Fender} R., 2003, \mnras, 345, L19

\bibitem[{{Martin} {et~al}\mbox{.}(2005){Martin}, {Fanson}, {Schiminovich},
  {Morrissey}, {Friedman}, {Barlow}, {Conrow}, {Grange}, {Jelinsky},
  {Milliard}, {Siegmund}, {Bianchi}, {Byun}, {Donas}, {Forster}, {Heckman},
  {Lee}, {Madore}, {Malina}, {Neff}, {Rich}, {Small}, {Surber}, {Szalay},
  {Welsh}, \& {Wyder}}]{2005ApJ...619L...1M}
{Martin} D.~C. {et~al.}, 2005, \apjl, 619, L1

\bibitem[{{Martin} {et~al}\mbox{.}(2007){Martin}, {Wyder}, {Schiminovich},
  {Barlow}, {Forster}, {Friedman}, {Morrissey}, {Neff}, {Seibert}, {Small},
  {Welsh}, {Bianchi}, {Donas}, {Heckman}, {Lee}, {Madore}, {Milliard}, {Rich},
  {Szalay}, \& {Yi}}]{2007ApJS..173..342M}
{Martin} D.~C. {et~al.}, 2007, \apjs, 173, 342

\bibitem[{{Masters} {et~al}\mbox{.}(2010{\natexlab{a}}){Masters}, {Mosleh},
  {Romer}, {Nichol}, {Bamford}, {Schawinski}, {Lintott}, {Andreescu},
  {Campbell}, {Crowcroft}, {Doyle}, {Edmondson}, {Murray}, {Raddick}, {Slosar},
  {Szalay}, \& {Vandenberg}}]{2010MNRAS.405..783M}
{Masters} K.~L. {et~al.}, 2010{\natexlab{a}}, \mnras, 405, 783

\bibitem[{{Masters} {et~al}\mbox{.}(2010{\natexlab{b}}){Masters}, {Nichol},
  {Bamford}, {Mosleh}, {Lintott}, {Andreescu}, {Edmondson}, {Keel}, {Murray},
  {Raddick}, {Schawinski}, {Slosar}, {Szalay}, {Thomas}, \&
  {Vandenberg}}]{2010MNRAS.404..792M}
{Masters} K.~L. {et~al.}, 2010{\natexlab{b}}, \mnras, 404, 792

\bibitem[{{Masters} {et~al}\mbox{.}(2012){Masters}, {Nichol}, {Haynes}, {Keel},
  {Lintott}, {Simmons}, {Skibba}, {Bamford}, {Giovanelli}, \&
  {Schawinski}}]{2012MNRAS.424.2180M}
{Masters} K.~L. {et~al.}, 2012, \mnras, 424, 2180

\bibitem[{{Masters} {et~al}\mbox{.}(2011){Masters}, {Nichol}, {Hoyle},
  {Lintott}, {Bamford}, {Edmondson}, {Fortson}, {Keel}, {Schawinski}, {Smith},
  \& {Thomas}}]{2011MNRAS.411.2026M}
{Masters} K.~L. {et~al.}, 2011, \mnras, 411, 2026

\bibitem[{{McHardy} {et~al}\mbox{.}(2006){McHardy}, {Koerding}, {Knigge},
  {Uttley}, \& {Fender}}]{2006Natur.444..730M}
{McHardy} I.~M., {Koerding} E., {Knigge} C., {Uttley} P., {Fender} R.~P., 2006,
  \nat, 444, 730

\bibitem[{{Melvin} {et~al}\mbox{.}(2014){Melvin}, {Masters}, {Lintott},
  {Nichol}, {Simmons}, {Bamford}, {Casteels}, {Cheung}, {Edmondson}, {Fortson},
  {Schawinski}, {Skibba}, {Smith}, \& {Willett}}]{2014MNRAS.tmp...97M}
{Melvin} T. {et~al.}, 2014, \mnras

\bibitem[{{Mendez} {et~al}\mbox{.}(2011){Mendez}, {Coil}, {Lotz}, {Salim},
  {Moustakas}, \& {Simard}}]{2011ApJ...736..110M}
{Mendez} A.~J., {Coil} A.~L., {Lotz} J., {Salim} S., {Moustakas} J., {Simard}
  L., 2011, \apj, 736, 110

\bibitem[{{Muzzin} {et~al}\mbox{.}(2013){Muzzin}, {Marchesini}, {Stefanon},
  {Franx}, {McCracken}, {Milvang-Jensen}, {Dunlop}, {Fynbo}, {Le Fevre},
  {Brammer}, \& {Labbe}}]{2013arXiv1303.4409M}
{Muzzin} A. {et~al.}, 2013, ArXiv e-prints, 1303.4409

\bibitem[{{Muzzin} {et~al}\mbox{.}(2012){Muzzin}, {Wilson}, {Yee}, {Gilbank},
  {Hoekstra}, {Demarco}, {Balogh}, {van Dokkum}, {Franx}, {Ellingson}, {Hicks},
  {Nantais}, {Noble}, {Lacy}, {Lidman}, {Rettura}, {Surace}, \&
  {Webb}}]{2012ApJ...746..188M}
{Muzzin} A. {et~al.}, 2012, \apj, 746, 188

\bibitem[{{Nandra} {et~al}\mbox{.}(2007){Nandra}, {Georgakakis}, {Willmer},
  {Cooper}, {Croton}, {Davis}, {Faber}, {Koo}, {Laird}, \&
  {Newman}}]{2007ApJ...660L..11N}
{Nandra} K. {et~al.}, 2007, \apjl, 660, L11

\bibitem[{{Noeske} {et~al}\mbox{.}(2007){Noeske}, {Weiner}, {Faber},
  {Papovich}, {Koo}, {Somerville}, {Bundy}, {Conselice}, {Newman},
  {Schiminovich}, {Le Floc'h}, {Coil}, {Rieke}, {Lotz}, {Primack}, {Barmby},
  {Cooper}, {Davis}, {Ellis}, {Fazio}, {Guhathakurta}, {Huang}, {Kassin},
  {Martin}, {Phillips}, {Rich}, {Small}, {Willmer}, \&
  {Wilson}}]{2007ApJ...660L..43N}
{Noeske} K.~G. {et~al.}, 2007, \apjl, 660, L43

\bibitem[{{Oh} {et~al}\mbox{.}(2011){Oh}, {Sarzi}, {Schawinski}, \&
  {Yi}}]{2011ApJS..195...13O}
{Oh} K., {Sarzi} M., {Schawinski} K., {Yi} S.~K., 2011, \apjs, 195, 13

\bibitem[{{Padmanabhan} {et~al}\mbox{.}(2008){Padmanabhan}, {Schlegel},
  {Finkbeiner}, {Barentine}, {Blanton}, {Brewington}, {Gunn}, {Harvanek},
  {Hogg}, {Ivezi{\'c}}, {Johnston}, {Kent}, {Kleinman}, {Knapp}, {Krzesinski},
  {Long}, {Neilsen}, {Nitta}, {Loomis}, {Lupton}, {Roweis}, {Snedden},
  {Strauss}, \& {Tucker}}]{2008ApJ...674.1217P}
{Padmanabhan} N. {et~al.}, 2008, \apj, 674, 1217

\bibitem[{{Pakull} {et~al}\mbox{.}(2010){Pakull}, {Soria}, \&
  {Motch}}]{2010Natur.466..209P}
{Pakull} M.~W., {Soria} R., {Motch} C., 2010, \nat, 466, 209

\bibitem[{{Peng} {et~al}\mbox{.}(2010){Peng}, {Lilly}, {Kova{\v c}},
  {Bolzonella}, {Pozzetti}, {Renzini}, {Zamorani}, {Ilbert}, {Knobel},
  {Iovino}, {Maier}, {Cucciati}, {Tasca}, {Carollo}, {Silverman}, {Kampczyk},
  {de Ravel}, {Sanders}, {Scoville}, {Contini}, {Mainieri}, {Scodeggio},
  {Kneib}, {Le F{\`e}vre}, {Bardelli}, {Bongiorno}, {Caputi}, {Coppa}, {de la
  Torre}, {Franzetti}, {Garilli}, {Lamareille}, {Le Borgne}, {Le Brun},
  {Mignoli}, {Perez Montero}, {Pello}, {Ricciardelli}, {Tanaka}, {Tresse},
  {Vergani}, {Welikala}, {Zucca}, {Oesch}, {Abbas}, {Barnes}, {Bordoloi},
  {Bottini}, {Cappi}, {Cassata}, {Cimatti}, {Fumana}, {Hasinger}, {Koekemoer},
  {Leauthaud}, {Maccagni}, {Marinoni}, {McCracken}, {Memeo}, {Meneux}, {Nair},
  {Porciani}, {Presotto}, \& {Scaramella}}]{2010ApJ...721..193P}
{Peng} Y.-j. {et~al.}, 2010, \apj, 721, 193

\bibitem[{{Salim} {et~al}\mbox{.}(2007){Salim}, {Rich}, {Charlot},
  {Brinchmann}, {Johnson}, {Schiminovich}, {Seibert}, {Mallery}, {Heckman},
  {Forster}, {Friedman}, {Martin}, {Morrissey}, {Neff}, {Small}, {Wyder},
  {Bianchi}, {Donas}, {Lee}, {Madore}, {Milliard}, {Szalay}, {Welsh}, \&
  {Yi}}]{2007ApJS..173..267S}
{Salim} S. {et~al.}, 2007, \apjs, 173, 267

\bibitem[{{Sarzi} {et~al}\mbox{.}(2006){Sarzi}, {Falc{\'o}n-Barroso}, {Davies},
  {Bacon}, {Bureau}, {Cappellari}, {de Zeeuw}, {Emsellem}, {Fathi},
  {Krajnovi{\'c}}, {Kuntschner}, {McDermid}, \&
  {Peletier}}]{2006MNRAS.366.1151S}
{Sarzi} M. {et~al.}, 2006, \mnras, 366, 1151

\bibitem[{{Schawinski}(2009)}]{2009AIPC.1201...17S}
{Schawinski} K., 2009, in American Institute of Physics Conference Series, Vol.
  1201, American Institute of Physics Conference Series, {Heinz} S., {Wilcots}
  E., eds., pp. 17--20

\bibitem[{{Schawinski}(2012)}]{2012arXiv1206.2661S}
{Schawinski} K., 2012, ArXiv e-prints, 1206.2661

\bibitem[{{Schawinski} {et~al}\mbox{.}(2010{\natexlab{a}}){Schawinski},
  {Dowlin}, {Thomas}, {Urry}, \& {Edmondson}}]{2010ApJ...714L.108S}
{Schawinski} K., {Dowlin} N., {Thomas} D., {Urry} C.~M., {Edmondson} E.,
  2010{\natexlab{a}}, \apjl, 714, L108

\bibitem[{{Schawinski} {et~al}\mbox{.}(2007{\natexlab{a}}){Schawinski},
  {Kaviraj}, {Khochfar}, {Yoon}, {Yi}, {Deharveng}, {Boselli}, {Barlow},
  {Conrow}, {Forster}, {Friedman}, {Martin}, {Morrissey}, {Neff},
  {Schiminovich}, {Seibert}, {Small}, {Wyder}, {Bianchi}, {Donas}, {Heckman},
  {Lee}, {Madore}, {Milliard}, {Rich}, \& {Szalay}}]{2007ApJS..173..512S}
{Schawinski} K. {et~al.}, 2007{\natexlab{a}}, \apjs, 173, 512

\bibitem[{{Schawinski} {et~al}\mbox{.}(2009{\natexlab{a}}){Schawinski},
  {Lintott}, {Thomas}, {Sarzi}, {Andreescu}, {Bamford}, {Kaviraj}, {Khochfar},
  {Land}, {Murray}, {Nichol}, {Raddick}, {Slosar}, {Szalay}, {Vandenberg}, \&
  {Yi}}]{2009MNRAS.396..818S}
{Schawinski} K. {et~al.}, 2009{\natexlab{a}}, \mnras, 396, 818

\bibitem[{{Schawinski} {et~al}\mbox{.}(2009{\natexlab{b}}){Schawinski},
  {Lintott}, {Thomas}, {Kaviraj}, {Viti}, {Silk}, {Maraston}, {Sarzi}, {Yi},
  {Joo}, {Daddi}, {Bayet}, {Bell}, \& {Zuntz}}]{2009ApJ...690.1672S}
{Schawinski} K. {et~al.}, 2009{\natexlab{b}}, \apj, 690, 1672

\bibitem[{{Schawinski} {et~al}\mbox{.}(2012){Schawinski}, {Simmons}, {Urry},
  {Treister}, \& {Glikman}}]{2012MNRAS.425L..61S}
{Schawinski} K., {Simmons} B.~D., {Urry} C.~M., {Treister} E., {Glikman} E.,
  2012, \mnras, 425, L61

\bibitem[{{Schawinski} {et~al}\mbox{.}(2007{\natexlab{b}}){Schawinski},
  {Thomas}, {Sarzi}, {Maraston}, {Kaviraj}, {Joo}, {Yi}, \&
  {Silk}}]{2007MNRAS.382.1415S}
{Schawinski} K., {Thomas} D., {Sarzi} M., {Maraston} C., {Kaviraj} S., {Joo}
  S.-J., {Yi} S.~K., {Silk} J., 2007{\natexlab{b}}, \mnras, 382, 1415

\bibitem[{{Schawinski} {et~al}\mbox{.}(2011){Schawinski}, {Treister}, {Urry},
  {Cardamone}, {Simmons}, \& {Yi}}]{2011ApJ...727L..31S}
{Schawinski} K., {Treister} E., {Urry} C.~M., {Cardamone} C.~N., {Simmons} B.,
  {Yi} S.~K., 2011, \apjl, 727, L31

\bibitem[{{Schawinski} {et~al}\mbox{.}(2010{\natexlab{b}}){Schawinski}, {Urry},
  {Virani}, {Coppi}, {Bamford}, {Treister}, {Lintott}, {Sarzi}, {Keel},
  {Kaviraj}, {Cardamone}, {Masters}, {Ross}, {Andreescu}, {Murray}, {Nichol},
  {Raddick}, {Slosar}, {Szalay}, {Thomas}, \&
  {Vandenberg}}]{2010ApJ...711..284S}
{Schawinski} K. {et~al.}, 2010{\natexlab{b}}, \apj, 711, 284

\bibitem[{{Schawinski} {et~al}\mbox{.}(2009{\natexlab{c}}){Schawinski},
  {Virani}, {Simmons}, {Urry}, {Treister}, {Kaviraj}, \&
  {Kushkuley}}]{2009ApJ...692L..19S}
{Schawinski} K., {Virani} S., {Simmons} B., {Urry} C.~M., {Treister} E.,
  {Kaviraj} S., {Kushkuley} B., 2009{\natexlab{c}}, \apjl, 692, L19

\bibitem[{{Schiminovich} {et~al}\mbox{.}(2007){Schiminovich}, {Wyder},
  {Martin}, {Johnson}, {Salim}, {Seibert}, {Treyer}, {Budav{\'a}ri}, {Hoopes},
  {Zamojski}, {Barlow}, {Forster}, {Friedman}, {Morrissey}, {Neff}, {Small},
  {Bianchi}, {Donas}, {Heckman}, {Lee}, {Madore}, {Milliard}, {Rich}, {Szalay},
  {Welsh}, \& {Yi}}]{2007ApJS..173..315S}
{Schiminovich} D. {et~al.}, 2007, \apjs, 173, 315

\bibitem[{{Schmidt}(1959)}]{1959ApJ...129..243S}
{Schmidt} M., 1959, \apj, 129, 243

\bibitem[{{Shapiro} {et~al}\mbox{.}(2010){Shapiro}, {Falc{\'o}n-Barroso}, {van
  de Ven}, {de Zeeuw}, {Sarzi}, {Bacon}, {Bolatto}, {Cappellari}, {Croton},
  {Davies}, {Emsellem}, {Fakhouri}, {Krajnovi{\'c}}, {Kuntschner}, {McDermid},
  {Peletier}, {van den Bosch}, \& {van der Wolk}}]{2010MNRAS.402.2140S}
{Shapiro} K.~L. {et~al.}, 2010, \mnras, 402, 2140

\bibitem[{{Sheth} {et~al}\mbox{.}(2008){Sheth}, {Elmegreen}, {Elmegreen},
  {Capak}, {Abraham}, {Athanassoula}, {Ellis}, {Mobasher}, {Salvato},
  {Schinnerer}, {Scoville}, {Spalsbury}, {Strubbe}, {Carollo}, {Rich}, \&
  {West}}]{2008ApJ...675.1141S}
{Sheth} K. {et~al.}, 2008, \apj, 675, 1141

\bibitem[{{Silverman} {et~al}\mbox{.}(2008){Silverman}, {Mainieri}, {Lehmer},
  {Alexander}, {Bauer}, {Bergeron}, {Brandt}, {Gilli}, {Hasinger}, {Schneider},
  {Tozzi}, {Vignali}, {Koekemoer}, {Miyaji}, {Popesso}, {Rosati}, \&
  {Szokoly}}]{2008ApJ...675.1025S}
{Silverman} J.~D. {et~al.}, 2008, \apj, 675, 1025

\bibitem[{{Simmons} {et~al}\mbox{.}(2012){Simmons}, {Urry}, {Schawinski},
  {Cardamone}, \& {Glikman}}]{2012ApJ...761...75S}
{Simmons} B.~D., {Urry} C.~M., {Schawinski} K., {Cardamone} C., {Glikman} E.,
  2012, \apj, 761, 75

\bibitem[{{Skibba} {et~al}\mbox{.}(2009){Skibba}, {Bamford}, {Nichol},
  {Lintott}, {Andreescu}, {Edmondson}, {Murray}, {Raddick}, {Schawinski},
  {Slosar}, {Szalay}, {Thomas}, \& {Vandenberg}}]{2009MNRAS.399..966S}
{Skibba} R.~A. {et~al.}, 2009, \mnras, 399, 966

\bibitem[{{Skibba} {et~al}\mbox{.}(2012){Skibba}, {Masters}, {Nichol},
  {Zehavi}, {Hoyle}, {Edmondson}, {Bamford}, {Cardamone}, {Keel}, {Lintott}, \&
  {Schawinski}}]{2012MNRAS.423.1485S}
{Skibba} R.~A. {et~al.}, 2012, \mnras, 423, 1485

\bibitem[{{Skrutskie} {et~al}\mbox{.}(2006){Skrutskie}, {Cutri}, {Stiening},
  {Weinberg}, {Schneider}, {Carpenter}, {Beichman}, {Capps}, {Chester},
  {Elias}, {Huchra}, {Liebert}, {Lonsdale}, {Monet}, {Price}, {Seitzer},
  {Jarrett}, {Kirkpatrick}, {Gizis}, {Howard}, {Evans}, {Fowler}, {Fullmer},
  {Hurt}, {Light}, {Kopan}, {Marsh}, {McCallon}, {Tam}, {Van Dyk}, \&
  {Wheelock}}]{2006AJ....131.1163S}
{Skrutskie} M.~F. {et~al.}, 2006, \aj, 131, 1163

\bibitem[{{Sodr{\'e}} {et~al}\mbox{.}(2013){Sodr{\'e}}, {Ribeiro da Silva}, \&
  {Santos}}]{2013arXiv1306.6552S}
{Sodr{\'e}}, Jr. L., {Ribeiro da Silva} A., {Santos} W.~A., 2013, ArXiv
  e-prints, 1306.6552

\bibitem[{{Springel} {et~al}\mbox{.}(2005){Springel}, {Di Matteo}, \&
  {Hernquist}}]{2005ApJ...620L..79S}
{Springel} V., {Di Matteo} T., {Hernquist} L., 2005, \apjl, 620, L79

\bibitem[{{Strateva} {et~al}\mbox{.}(2001){Strateva}, {Ivezi{\'c}}, {Knapp},
  {Narayanan}, {Strauss}, {Gunn}, {Lupton}, {Schlegel}, {Bahcall}, {Brinkmann},
  {Brunner}, {Budav{\'a}ri}, {Csabai}, {Castander}, {Doi}, {Fukugita}, {Gy{\H
  o}ry}, {Hamabe}, {Hennessy}, {Ichikawa}, {Kunszt}, {Lamb}, {McKay},
  {Okamura}, {Racusin}, {Sekiguchi}, {Schneider}, {Shimasaku}, \&
  {York}}]{2001AJ....122.1861S}
{Strateva} I. {et~al.}, 2001, \aj, 122, 1861

\bibitem[{{Taylor}(2011)}]{2011ascl.soft01010T}
{Taylor} M., 2011, Astrophysics Source Code Library, 1010

\bibitem[{{Taylor}(2005)}]{2005ASPC..347...29T}
{Taylor} M.~B., 2005, in Astronomical Society of the Pacific Conference Series,
  Vol. 347, Astronomical Data Analysis Software and Systems XIV, {Shopbell} P.,
  {Britton} M., {Ebert} R., eds., p.~29

\bibitem[{{Teng} {et~al}\mbox{.}(2012){Teng}, {Schawinski}, {Urry}, {Darg},
  {Kaviraj}, {Oh}, {Bonning}, {Cardamone}, {Keel}, {Lintott}, {Simmons}, \&
  {Treister}}]{2012ApJ...753..165T}
{Teng} S.~H. {et~al.}, 2012, \apj, 753, 165

\bibitem[{{Tinsley}(1968)}]{1968ApJ...151..547T}
{Tinsley} B.~M., 1968, \apj, 151, 547

\bibitem[{{Tinsley} \& {Gunn}(1976)}]{1976ApJ...203...52T}
{Tinsley} B.~M., {Gunn} J.~E., 1976, \apj, 203, 52

\bibitem[{{Tinsley} \& {Larson}(1978)}]{1978ApJ...221..554T}
{Tinsley} B.~M., {Larson} R.~B., 1978, \apj, 221, 554

\bibitem[{{Tonnesen}(2011)}]{2011nha..confE...8T}
{Tonnesen} S., 2011, in New Horizons in Astronomy, Proceedings of the Frank N.
  Bash Symposium 2011, held October 9-11, 2011. Austin, Texas, USA. Edited by
  S. Salviander, J. Green, and A. Pawlik. Published online at <A
  href=''http://pos.sissa.it/cgi-bin/reader/conf.cgi?confid=149''>http://pos.sissa.it/cgi-bin/reader/conf.cgi?confid=149</A>.,
  id.8

\bibitem[{{Treister} {et~al}\mbox{.}(2012){Treister}, {Schawinski}, {Urry}, \&
  {Simmons}}]{2012ApJ...758L..39T}
{Treister} E., {Schawinski} K., {Urry} C.~M., {Simmons} B.~D., 2012, \apjl,
  758, L39

\bibitem[{{van den Bergh}(1976)}]{1976ApJ...206..883V}
{van den Bergh} S., 1976, \apj, 206, 883

\bibitem[{{Vollmer}(2013)}]{2013pss6.book..207V}
{Vollmer} B., 2013, {The Influence of Environment on Galaxy Evolution},
  {Oswalt} T.~D., {Keel} W.~C., eds., p. 207

\bibitem[{{Wetzel} {et~al}\mbox{.}(2013){Wetzel}, {Tinker}, {Conroy}, \& {van
  den Bosch}}]{2013MNRAS.432..336W}
{Wetzel} A.~R., {Tinker} J.~L., {Conroy} C., {van den Bosch} F.~C., 2013,
  \mnras, 432, 336

\bibitem[{{Wild} {et~al}\mbox{.}(2007){Wild}, {Kauffmann}, {Heckman},
  {Charlot}, {Lemson}, {Brinchmann}, {Reichard}, \&
  {Pasquali}}]{2007MNRAS.381..543W}
{Wild} V., {Kauffmann} G., {Heckman} T., {Charlot} S., {Lemson} G.,
  {Brinchmann} J., {Reichard} T., {Pasquali} A., 2007, \mnras, 381, 543

\bibitem[{{Williams} {et~al}\mbox{.}(2009){Williams}, {Quadri}, {Franx}, {van
  Dokkum}, \& {Labb{\'e}}}]{2009ApJ...691.1879W}
{Williams} R.~J., {Quadri} R.~F., {Franx} M., {van Dokkum} P., {Labb{\'e}} I.,
  2009, \apj, 691, 1879

\bibitem[{{Wolf} {et~al}\mbox{.}(2009){Wolf}, {Arag{\'o}n-Salamanca}, {Balogh},
  {Barden}, {Bell}, {Gray}, {Peng}, {Bacon}, {Barazza}, {B{\"o}hm}, {Caldwell},
  {Gallazzi}, {H{\"a}u{\ss}ler}, {Heymans}, {Jahnke}, {Jogee}, {van Kampen},
  {Lane}, {McIntosh}, {Meisenheimer}, {Papovich}, {S{\'a}nchez}, {Taylor},
  {Wisotzki}, \& {Zheng}}]{2009MNRAS.393.1302W}
{Wolf} C. {et~al.}, 2009, \mnras, 393, 1302

\bibitem[{{Wong} {et~al}\mbox{.}(2012){Wong}, {Schawinski}, {Kaviraj},
  {Masters}, {Nichol}, {Lintott}, {Keel}, {Darg}, {Bamford}, {Andreescu},
  {Murray}, {Raddick}, {Szalay}, {Thomas}, \&
  {Vandenberg}}]{2012MNRAS.420.1684W}
{Wong} O.~I. {et~al.}, 2012, \mnras, 420, 1684

\bibitem[{{Wyder} {et~al}\mbox{.}(2007){Wyder}, {Martin}, {Schiminovich},
  {Seibert}, {Budav{\'a}ri}, {Treyer}, {Barlow}, {Forster}, {Friedman},
  {Morrissey}, {Neff}, {Small}, {Bianchi}, {Donas}, {Heckman}, {Lee}, {Madore},
  {Milliard}, {Rich}, {Szalay}, {Welsh}, \& {Yi}}]{2007ApJS..173..293W}
{Wyder} T.~K. {et~al.}, 2007, \apjs, 173, 293

\bibitem[{{Yang} {et~al}\mbox{.}(2007){Yang}, {Mo}, {van den Bosch},
  {Pasquali}, {Li}, \& {Barden}}]{2007ApJ...671..153Y}
{Yang} X., {Mo} H.~J., {van den Bosch} F.~C., {Pasquali} A., {Li} C., {Barden}
  M., 2007, \apj, 671, 153

\bibitem[{{Yi} {et~al}\mbox{.}(2005){Yi}, {Yoon}, {Kaviraj}, {Deharveng},
  {Rich}, {Salim}, {Boselli}, {Lee}, {Ree}, {Sohn}, {Rey}, {Lee}, {Rhee},
  {Bianchi}, {Byun}, {Donas}, {Friedman}, {Heckman}, {Jelinsky}, {Madore},
  {Malina}, {Martin}, {Milliard}, {Morrissey}, {Neff}, {Schiminovich},
  {Siegmund}, {Small}, {Szalay}, {Jee}, {Kim}, {Barlow}, {Forster}, {Welsh}, \&
  {Wyder}}]{2005ApJ...619L.111Y}
{Yi} S.~K. {et~al.}, 2005, \apjl, 619, L111

\bibitem[{{York} {et~al}\mbox{.}(2000){York}, {Adelman}, {Anderson},
  {Anderson}, {Annis}, {Bahcall}, {Bakken}, {Barkhouser}, {Bastian}, {Berman},
  {Boroski}, {Bracker}, {Briegel}, {Briggs}, {Brinkmann}, {Brunner}, {Burles},
  {Carey}, {Carr}, {Castander}, {Chen}, {Colestock}, {Connolly}, {Crocker},
  {Csabai}, {Czarapata}, {Davis}, {Doi}, {Dombeck}, {Eisenstein}, {Ellman},
  {Elms}, {Evans}, {Fan}, {Federwitz}, {Fiscelli}, {Friedman}, {Frieman},
  {Fukugita}, {Gillespie}, {Gunn}, {Gurbani}, {de Haas}, {Haldeman}, {Harris},
  {Hayes}, {Heckman}, {Hennessy}, {Hindsley}, {Holm}, {Holmgren}, {Huang},
  {Hull}, {Husby}, {Ichikawa}, {Ichikawa}, {Ivezi{\'c}}, {Kent}, {Kim},
  {Kinney}, {Klaene}, {Kleinman}, {Kleinman}, {Knapp}, {Korienek}, {Kron},
  {Kunszt}, {Lamb}, {Lee}, {Leger}, {Limmongkol}, {Lindenmeyer}, {Long},
  {Loomis}, {Loveday}, {Lucinio}, {Lupton}, {MacKinnon}, {Mannery}, {Mantsch},
  {Margon}, {McGehee}, {McKay}, {Meiksin}, {Merelli}, {Monet}, {Munn},
  {Narayanan}, {Nash}, {Neilsen}, {Neswold}, {Newberg}, {Nichol}, {Nicinski},
  {Nonino}, {Okada}, {Okamura}, {Ostriker}, {Owen}, {Pauls}, {Peoples},
  {Peterson}, {Petravick}, {Pier}, {Pope}, {Pordes}, {Prosapio},
  {Rechenmacher}, {Quinn}, {Richards}, {Richmond}, {Rivetta}, {Rockosi},
  {Ruthmansdorfer}, {Sandford}, {Schlegel}, {Schneider}, {Sekiguchi}, {Sergey},
  {Shimasaku}, {Siegmund}, {Smee}, {Smith}, {Snedden}, {Stone}, {Stoughton},
  {Strauss}, {Stubbs}, {SubbaRao}, {Szalay}, {Szapudi}, {Szokoly}, {Thakar},
  {Tremonti}, {Tucker}, {Uomoto}, {Vanden Berk}, {Vogeley}, {Waddell}, {Wang},
  {Watanabe}, {Weinberg}, {Yanny}, \& {Yasuda}}]{2000AJ....120.1579Y}
{York} D.~G. {et~al.}, 2000, \aj, 120, 1579

\end{thebibliography}

\bsp

\newpage
\appendix
\section{Does the early star formation history of main sequence galaxies have an appreciable effect on post-quenching evolution?}
\label{appendix_a}

%-----------------------------------------------------------------------------------------------------------------------------------
\begin{figure*}
\begin{center}

\includegraphics[width=0.49\textwidth]{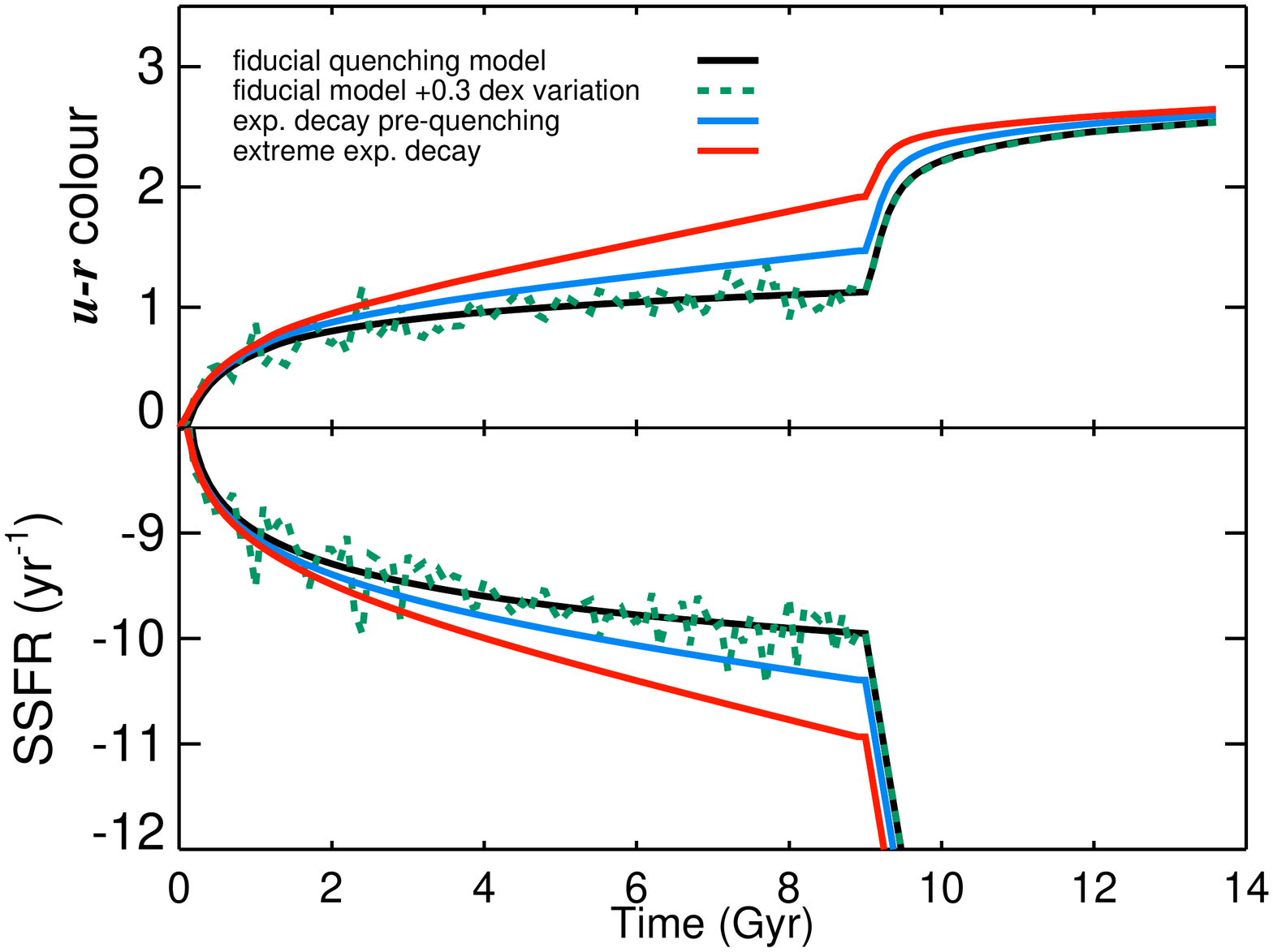}
\includegraphics[width=0.49\textwidth]{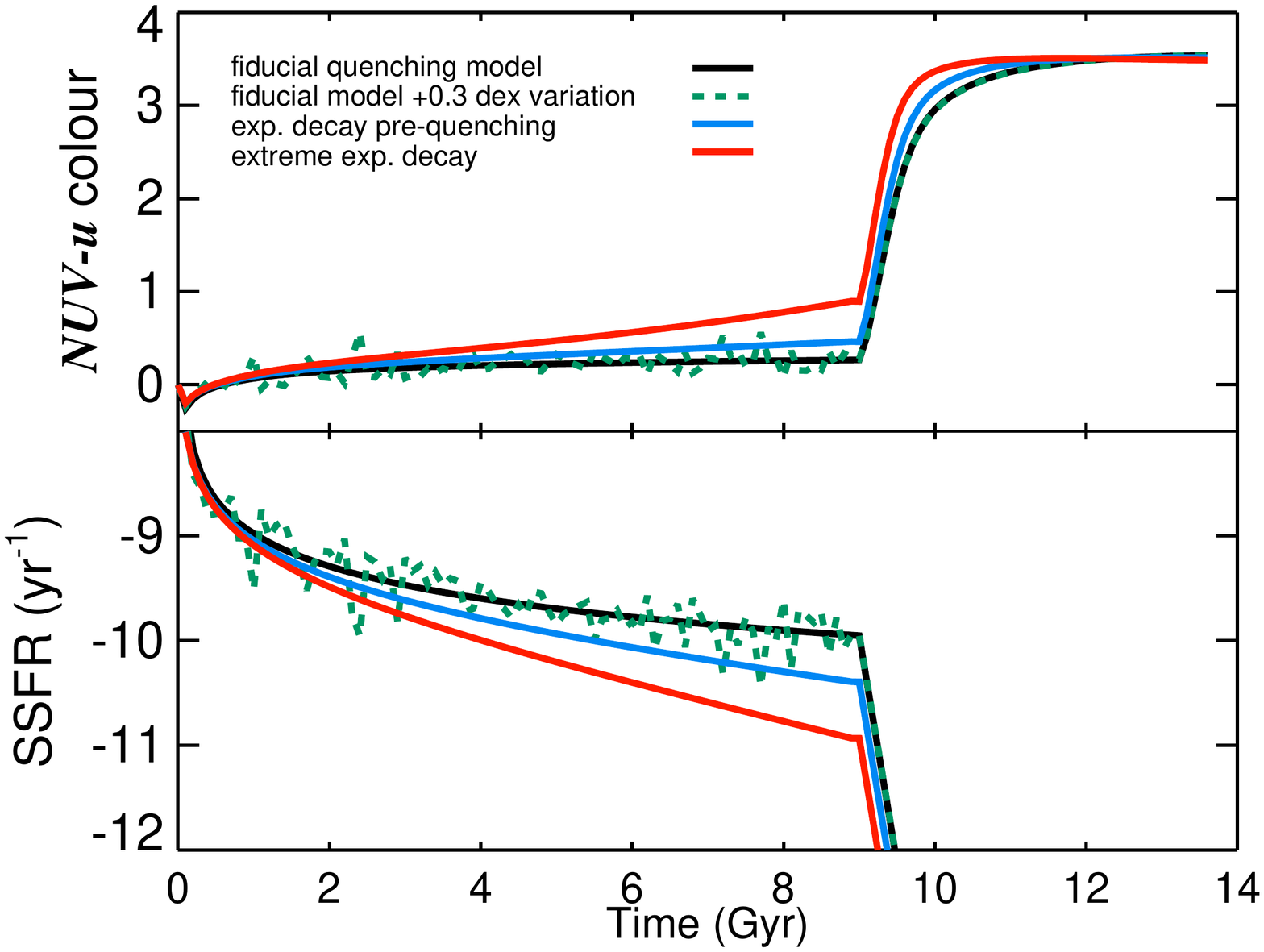}

\caption{Evolution of $u-r$ colour (\textit{left}) and $NUV-r$ colour (\textit{right}) as a function of time for a set of four models. Below the colour evolution of each model, we show the specific star formation rate as a function of time. Model 1 (black line): fiducial model of constant SFR, followed by exponential decay, in this case with $\tau_{\rm quench} = 100$ Myr. Model 2 (green dashed line): same as Model 1, but with random perturbations prior to quenching. Model 3 (light blue line): rather than a constant SFR, the pre-quenching star formation history is a gently exponentially declining SFR. This SFR builds up a more significant old stellar population by the time of quenching. The model is tuned to have an SSFR at the quenching time $t_{\rm quench} = 9$ Gyr that is at the lower edge of the Main Sequence. Model 4 (red line): same as Model 3, but with a more strongly exponentially declining SFR prior to quenching. This Model has an SSFR $\sim$1 dex below the Main Sequence at $t_{\rm quench}$. Due to its low SSFR at $t_{\rm quench}$, Model 4 is incompatible with the Main Sequence. The other models, do not show significant offsets from the fiducial quenching model as they all have the following salient features: at $t_{\rm quench}$, they have blue cloud $u-r$ and $NUV-u$ colours, and after the quenching event, they become redder at essentially the same rate. Thus, all SFHs which place a galaxy on the main sequence at the point of quenching yield similar quenching timescale. }

\label{fig:color_time}

\end{center}
\end{figure*}
%-----------------------------------------------------------------------------------------------------------------------------------

We use the $NUVur$ diagram (Figure \ref{fig:evol1}) to constrain the quenching time scales of galaxies using a simple model star formation history: constant SFR for 9 Gyr, followed by an exponential decline of varying time scale $\tau$. The model assumes that all star-forming galaxies are on the main sequence prior to the quenching event, as is appropriate to our goal of describing the bulk of normal galaxies. But do all galaxies on the main sequence have comparable colours; that is, are the starting points of all star forming galaxies comparable? In order to test this, we plot the colour evolution of $u-r$ and $NUV-u$ in Figure \ref{fig:color_time}. We explore a range of models: 

\begin{enumerate}
\item Model 1 (black line): fiducial model of constant SFR, followed by exponential decay, in this case with $\tau_{\rm quench} = 100$ Myr. 
\item Model 2 (green dashed line): same as Model 1, but with Gaussian random perturbations in the amplitude prior to quenching with a dispersion of a factor 2.
\item Model 3 (light blue line): rather than a constant SFR, the pre-quenching star formation history is a gently exponentially declining SFR ($\tau = 5.0$ Gyr), which builds up a more significant old stellar population by the time of quenching. The model is tuned to have an SSFR at the quenching time $t_{\rm quench} = 9$ Gyr that is at the lower edge of the Main Sequence. 
\item Model 4 (red line): same as Model 3, but with a more strongly exponentially declining SFR prior to quenching  ($\tau = 2.5$ Gyr. This Model has an SSFR ~1 dex below the Main Sequence at $t_{\rm quench}$, and thus is (just a bit) outside the suite of galaxies whose quenching is discussed in this paper. 
\end{enumerate}

These model tracks demonstrate that a wide range of initial star-formation histories create galaxies with similar main sequence $u-r$ and $NUV-u$ colours at the point of quenching. This is because $NUV-u$ and $u-r$ colours are dominated by current and recent recent stellar populations. Thus, star formation histories that place a galaxy on the main sequence will also result in similarly blue colours. Once quenching sets in, the movement towards the red is similarly governed by young stellar populations and and thus the time scales are similar regardless of the pre-quenching star formation history.

We combine these diagrams and place the same four model tracks on a version of Figure \ref{fig:evol1} in Figure \ref{fig:color_check}. We show how those four model tracks evolve: Models 1, 2 and 3 are virtually on top of each other and indicate a short quenching time scale, as expected from the assumed $\tau_{\rm quench} = 100$ Myr. Model 4 (red) -- which is ruled out due to its low SSFR -- is offset and would indicate somewhat longer time scales, but is still not comparable to the "slow" quenching tracks which best fit late-type galaxies. We  conclude that the time scales deduced from the $NUVur$ diagram are robust with respect to initial star formation histories because the main sequence dictates similar colours at the point of quenching.

%-----------------------------------------------------------------------------------------------------------------------------------
\begin{figure}
\begin{center}

\includegraphics[width=0.49\textwidth]{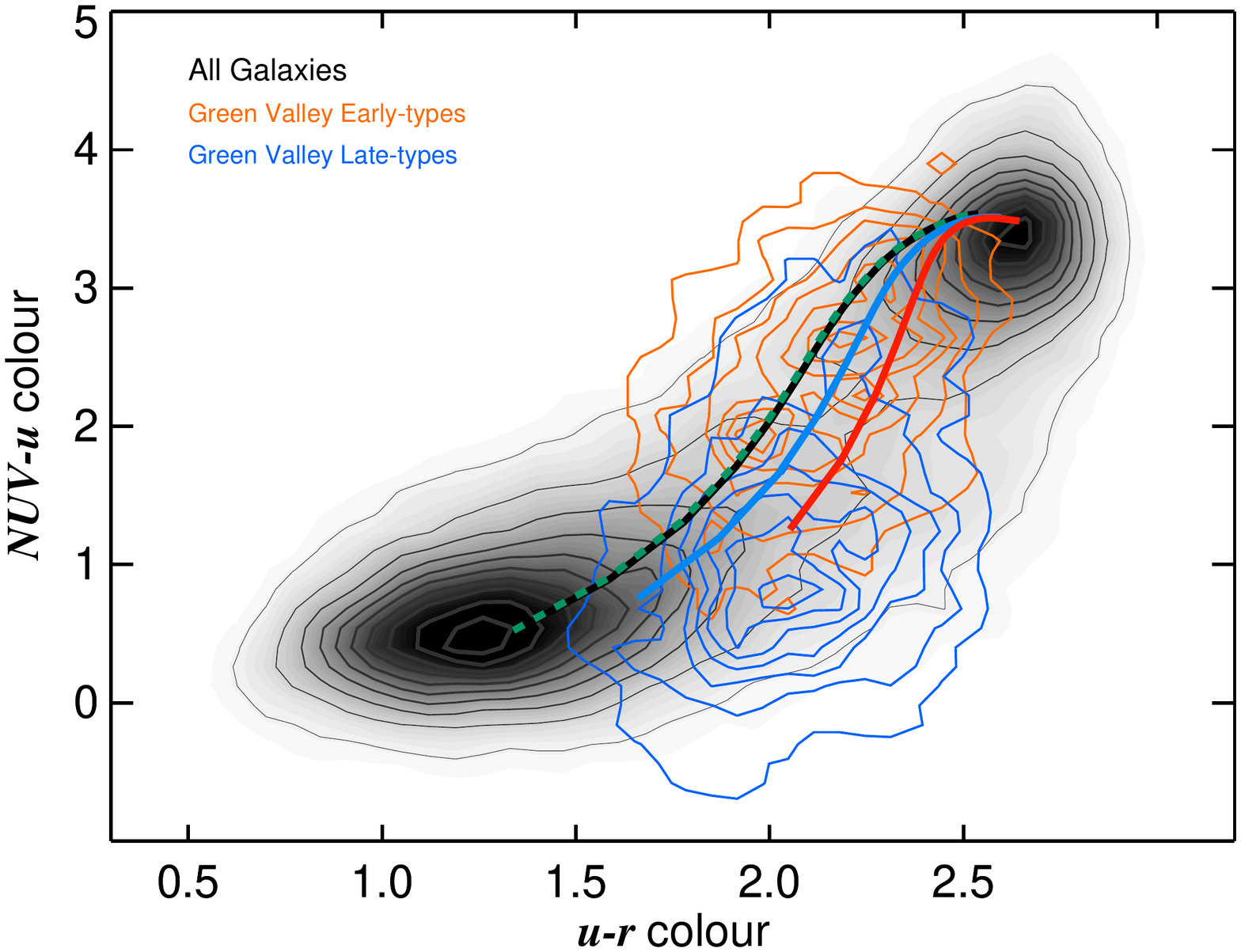}

\caption{The $NUVur$ colour diagram, similar to Figure \ref{fig:evol1}, with the same Model tracks shown as on Figure \ref{fig:color_time}. All models that are on the main sequence at the time of quenching (black, green, light blue) yield a short quenching. The red model, which is incompatible with the Main Sequence due to having a low SSFR at the quenching time, is somewhat offset, but still yields a short quenching time scale on the $NUVur$ diagram. Together with Figure \ref{fig:color_time}, this Figure illustrates the robustness of the $NUVur$ diagram as a quenching time and that all reasonable star formation histories that end up on the main sequence are blue, and reasonable starting points for our quenching model.}

\label{fig:color_check}

\end{center}
\end{figure}
%-----------------------------------------------------------------------------------------------------------------------------------

\label{lastpage}

\end{document}